\documentclass[aps,twocolumn,superscriptaddress,showpacs,showkeys,preprintnumbers,amsmath,amssymb,amsfonts,nofootinbib]{revtex4}

\pdfoutput=1

\usepackage{graphicx}
\usepackage{color}
\usepackage{slashed}
\usepackage{tabularx}
\usepackage{hyperref}
\usepackage{bbold}
\usepackage{bbm}
\usepackage{amsmath}
\usepackage{placeins}
\usepackage[normalem]{ulem}

\newcommand{\tr}{\mathrm{tr}}
\newcommand{\Tr}{\mathrm{Tr}}
\newcommand{\STr}{\mathrm{STr}}

\newcommand{\Nf}{N_{\text{f}}}
\newcommand{\Nc}{N_{\text{c}}}

\renewcommand{\Re}{{\text{Re}\,}}
\renewcommand{\Im}{{\text{Im}\,}}

\newcolumntype{L}[1]{>{\raggedright\arraybackslash}p{#1}} 
\newcolumntype{C}[1]{>{\centering\arraybackslash}p{#1}} 
\newcolumntype{R}[1]{>{\raggedleft\arraybackslash}p{#1}} 

\graphicspath{{.}{figures/}{mma/}}

\begin{document}

\title{Fermionic excitations at finite temperature and density}

\newcommand{\ECT}{European Centre for Theoretical Studies in Nuclear Physics and related Areas (ECT*) and Fondazione Bruno Kessler, Villa Tambosi, Strada delle Tabarelle 286, I-38123 Villazzano (TN), Italy}
\newcommand{\JLU}{Institut f\"ur Theoretische Physik, Justus-Liebig-Universit\"at Giessen, Heinrich-Buff-Ring 16, 35392 Giessen, Germany}
\newcommand{\TUDa}{Institut f\"ur Kernphysik (Theoriezentrum), Technische Universit\"at Darmstadt, Schlossgartenstr. 2, 64289 Darmstadt, Germany}
\newcommand{\GU}{Institute for Theoretical Physics, Goethe University, Max-von-Laue-Str.~1, D-60438 Frankfurt am Main, Germany}

\author{Ralf-Arno Tripolt}\affiliation{\GU}
\author{Dirk H.~Rischke}\affiliation{\GU}
\author{Lorenz von Smekal}\affiliation{\JLU}
\author{Jochen Wambach}\affiliation{\ECT}\affiliation{\TUDa}

\begin{abstract}
We study fermionic excitations in a hot and dense strongly interacting medium consisting of quarks and (pseudo-)scalar mesons. In particular, we use the two-flavor quark-meson model in combination with the Functional Renormalization Group approach, which allows to take into account the effects from thermal and quantum fluctuations. The resulting fermionic excitation spectrum is investigated by calculating the quark spectral function at finite temperature, quark chemical potential, and spatial momentum. This involves an analytic continuation from imaginary to real energies by extending the previously introduced analytically-continued FRG (aFRG) method to the present case. We identify three different collective excitations in the medium: the ordinary thermal quark, the plasmino mode, and an ultra-soft ``phonino'' mode. The dispersion relations of these modes are extracted from the quark spectral function. When compared to corresponding results from an FRG-improved one-loop calculation remarkable agreement is found.
\end{abstract}

\pacs{12.38.Aw, 12.38.Lg, 11.30.Rd}
\keywords{spectral function, analytic continuation, QCD, chiral symmetry}


\maketitle

\section{Introduction}\label{sec:introduction}

The spectrum of fermionic excitations in a hot relativistic plasma composed of quarks, antiquarks, and gluons (or electrons, positrons, and photons) exhibits interesting collective modes that have been studied for several decades by now \cite{Klimov1981, Klimov1982, Weldon1982, Weldon1983, Weldon1989, Pisarski1989, LebedevSmilga1990, BraatenPisarski1990, FrenkelTaylor1990, Petitgirard1992, BaymBlaizotSvetitsky1992,BlaizotOllitrault1993, VijaThoma1995, Lebellac:1996, Weldon2000,KitazawaKunihiroNemoto2006, KitazawaKunihiroNemoto2007, KitazawaKunihiroMitsutaniEtAl2008, KarschKitazawa2009, Kaczmarek:2012mb, HidakaSatowKunihiro2012, BlaizotSatow2014, KitazawaKunihiroNemoto2014, SuTywoniuk2015}. Various methods have been developed to study these systems, based on perturbation theory, i.e., the hard thermal loop (HTL) or hard dense loop (HDL) approximation \cite{Weldon1982, FrenkelTaylor1990, BraatenPisarski1990, VijaThoma1995}, as well as kinetic theory \cite{BlaizotIancu1993,BlaizotIancu1993a,BlaizotIancu1994}. One generally finds the emergence of collective phenomena on a ``soft'' momentum scale $p\sim gT$ or $p\sim g\mu$, where $g$ is the coupling constant, $T$ the temperature, and $\mu$ the fermion chemical potential. Already standard HTL calculations show the existence of two branches, namely the ordinary fermionic quasi-particle branch and the ``plasmino'' or hole mode, i.e., a collective excitation termed in analogy to the plasmon oscillation. Beyond-HTL calculations at finite temperature also reveal another collective excitation in the region of ultra-soft momenta, $p\sim g^2T,g^2 \mu$, which we will refer to as a ``phonino"
\cite{Kapusta1982,LebedevSmilga1990, KitazawaKunihiroNemoto2006, KitazawaKunihiroNemoto2007,HaradaNemoto2008,KitazawaKunihiroMitsutaniEtAl2008, HaradaNemotoYoshimoto2008, QinChangLiuEtAl2011, NakkagawaYokotaYoshida2012,   SatowHidakaKunihiro2011,HidakaSatowKunihiro2012, NakkagawaYokotaYoshida2012a, Satow2013, BlaizotSatow2014, KitazawaKunihiroNemoto2014, SuTywoniuk2015}.

In this work we investigate the existence and the properties of these fermionic excitations at finite temperature and density in the presence of fluctuations, thus going beyond the usual HTL and one-loop calculations. In order to include the effects from quantum and thermal fluctuations, we use the Functional Renormalization Group (FRG) approach and apply it to the quark-meson model as a low-energy model for the chiral aspects of two-flavor Quantum Chromodynamics (QCD). The resulting fermionic excitation spectrum is investigated by calculating the quark spectral function at finite temperature, quark chemical potential, and spatial momentum. This involves overcoming the analytic-continuation problem, i.e., the difficulty to extract real-time quantities such as spectral functions from a Euclidean (imaginary-time) framework like the FRG, in its usual formulation. We deal with this problem by using the analytically-continued FRG (aFRG) method, which was proposed in Refs.\ \cite{Kamikado2014,Tripolt2014, Tripolt2014a}. The aFRG method avoids the need for numerical reconstruction schemes, see for example Refs.\ \cite{Vidberg:1977,Jarrell:1996,Asakawa:2000tr,Qin:2013ufa, Dudal2013,FischerPawlowskiRothkopfEtAl2017,CyrolMitterPawlowskiEtAl2017, TripoltGublerUlybysheEtAl2018, Cyrol:2018xeq}, and exhibits a number of particular advantages: it preserves the underlying symmetry structures and their breaking patterns, and it is thermodynamically consistent in that the thermodynamic potential and the spectral functions are calculated on the same footing. The aFRG method has been successfully applied in different situations, for example to calculate mesonic in-medium spectral functions in Refs.\ \cite{Tripolt2014, Tripolt2014a, TripoltSmekalWambach2017}, vector- and axial-vector meson spectral functions in Refs.\ \cite{JungRenneckeTripoltEtAl2017}, and quark spectral functions at $T=0$ and $\mu=0$ in Ref.\ \cite{TripoltWeyrichSmekalEtAl2018}. The present work therefore is an extension of Ref.\ \cite{TripoltWeyrichSmekalEtAl2018} to finite temperature and chemical potential as well as finite spatial momentum relative to the heat bath.

The paper is organized as follows. In Sec.~\ref{sec:FRG} we briefly introduce the FRG framework and its application to the quark-meson model, and then apply the aFRG method to derive the flow equation for the real-time quark two-point function, from which the quark spectral functions are obtained. In Sec.~\ref{sec:1_loop} we discuss an FRG-improved one-loop calculation, which is used for comparison to the FRG computation. In Sec.~\ref{sec:results} our results for the quark spectral functions at finite temperature, finite quark-chemical potential, and finite spatial momentum as well as for the dispersion relations of the identified fermionic excitations are presented. We close with a summary and outlook in Sec.~\ref{sec:summary}. Further details are deferred to several appendices.

\section{Quark spectral function with the FRG}
\label{sec:FRG}

\subsection{FRG approach to the quark-meson model}
\label{sec:FRG_QM}

The FRG is a powerful non-perturbative approach with a wide range of applications, see e.g.~Refs.\ \cite{Berges:2000ew,Polonyi:2001se,Pawlowski:2005xe,Schaefer:2006sr,Kopietz2010,Braun:2011pp, Friman:2011zz, Gies2012} for reviews. It is usually formulated in (continuous) Euclidean space-time and combines Wilson's idea of the renormalization group in momentum space \cite{Wilson1971, Wilson1974} with functional methods in quantum field theory.

In the following we will use the formulation pioneered by Wetterich \cite{Wetterich:1992yh}, which aims at calculating the effective average action $\Gamma_k$, where $k$ is the renormalization-group scale. After choosing a suitable ansatz for the effective average action at the ultraviolet (UV) scale $k=\Lambda$, the effects of quantum and thermal fluctuations are gradually included until the full effective action $\Gamma=\Gamma_{k=0}$ is obtained in the limit $k\rightarrow0$.
The scale dependence of $\Gamma_k$ is given by the following flow equation,
\begin{align}
\label{eq:Wetterich}
\partial_k \Gamma_k[\phi,\psi,\bar{\psi}]=
&\frac{1}{2}\,\STr \left[\partial_kR_k \left(\Gamma_{k}^{(2)}[\phi,\psi,\bar{\psi}]+R_k\right)^{-1}\right],
\end{align}
where $R_k$ is a regulator function that suppresses momentum modes with momenta smaller than $k$, $\Gamma_{k}^{(2)}$ is the second functional derivative with respect to the fields, and the supertrace runs over field space, all internal indices, and also includes an integration over internal momenta. 

Our ansatz for the effective average action is based on the quark-meson model as a low-energy effective theory for the chiral aspects of QCD with two flavors \cite{Jungnickel:1995fp,Schaefer:2004en} and reads
\begin{align}\label{eq:gamma}
\Gamma_{k}[\phi, \psi, \bar\psi]=
\int d^{4}x \:\Big\{
&\bar{\psi}\left(\gamma_\mu\partial^\mu+
h(\sigma+i\vec{\tau}\cdot\vec{\pi}\gamma^{5}) -\mu \gamma_0 \right)\psi\nonumber\\
&+\frac{1}{2} (\partial_{\mu}\phi)^{2}+U_{k}(\phi^2)-c\sigma
\Big\},
\end{align}
with $\phi^2=\sigma^2+\vec\pi^2$, an effective potential $U_{k}(\phi^2)$, the quark chemical potential $\mu$, and $c\sigma$ an explicit chiral symmetry breaking term which plays the role of the (up/down) current quark mass in QCD. This ansatz represents the leading order in a derivative expansion, also called local potential approximation (LPA) \cite{Litim2001,Braun:2009si}. When inserting this ansatz into the Wetterich equation, one obtains the flow equation for the effective potential,
\begin{align}
\label{eq:flow_pot}
\partial_k U_k =\frac{k^3}{6\pi^2}\left( \frac{1}{2}I_{k,\sigma} + \frac{3}{2} I_{k,\pi} -2\Nc \Nf I_{k,\psi}\right),
\end{align}
where explicit expressions for the threshold functions $I_k$ are given in App.~\ref{app:FRG}. At the UV scale $\Lambda$ we choose the effective potential to be symmetric,
\begin{equation}
\label{eq:pot_UV}
U_\Lambda(\phi^{2}) =
\frac{1}{2}m_\Lambda^{2}\phi^{2} +
\frac{1}{4}\lambda_\Lambda(\phi^{2})^{2},
\end{equation}
and then solve the corresponding flow equation numerically using the so-called ``grid method'', see for example 
Ref.\ \cite{Schaefer:2004en}. This method is based on a finite-difference scheme to solve the FRG flow
equation. In case of discontinuities, such schemes do not always converge to the correct solution, and one should rather use finite-element or finite-volume methods, see Ref.\ \cite{GrossiWink2019}. However, as shown in
Ref.\ \cite{GrossiWink2019}, such discontinuities typically do not influence the minimum of the effective potential (i.e., the physical point), so that for the investigations presented in this paper the grid method is still expected to produce reliable results. In fact, we also checked explicitly that our results do not depend on the particular numerical method used by comparing them to results obtained from the so-called KT method \cite{KT2000}, i.e., a finite-volume method. The numerical values for the parameters are chosen as in Ref.\ \cite{TripoltWeyrichSmekalEtAl2018}, which yields the following physical values for the pion decay constant and the particle masses in the IR: $f_\pi=93.5$~MeV, $m_\pi=138$~MeV, $m_\sigma=509$~MeV, and $m_\psi=299$~MeV. The UV cutoff is chosen to be $\Lambda=1$~GeV. The solution for the scale-dependent effective potential is then used as input for the calculation of the quark two-point function and the quark spectral function.

\subsection{Quark two-point function and analytic continuation}
\label{sec:continuation}

\begin{figure*}[t!]
	\includegraphics[width=0.9\textwidth]{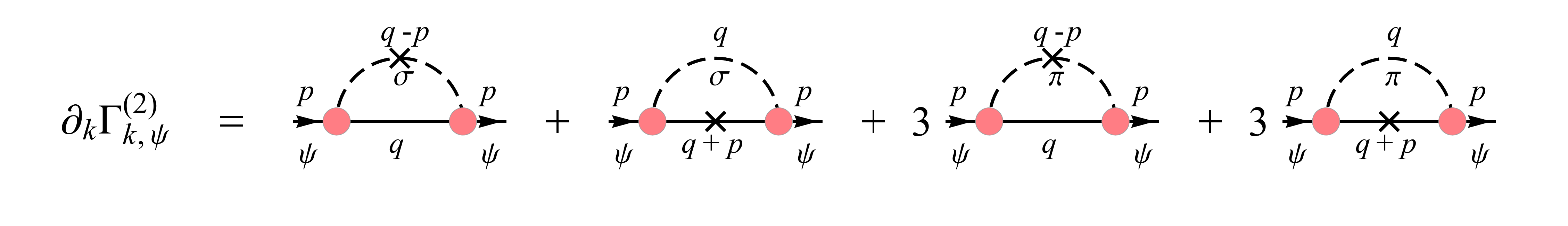}\\[-5mm]
	\includegraphics[width=0.9\textwidth]{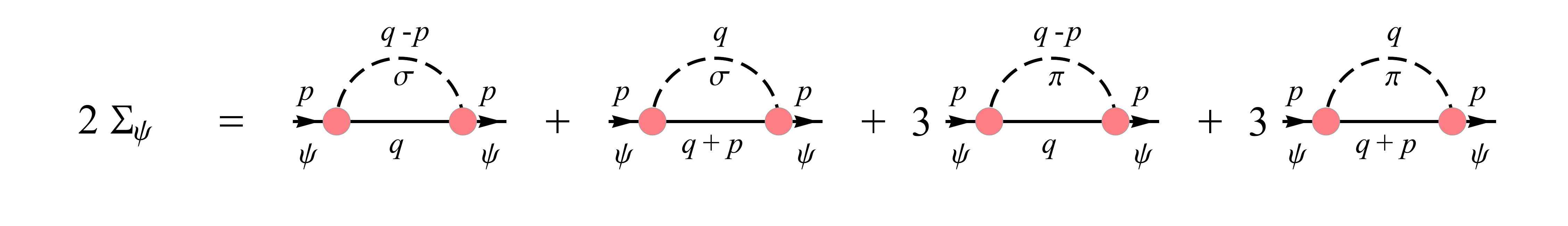}\vspace*{-5mm}
	\caption{Diagrammatic representation of the flow equation for the quark two-point function (top), cf.~Eq.~(\ref{eq:flow_gamma2}), and of the one-loop expression for the quark self-energy (bottom), cf.~Eq.~(\ref{eq:self_energy}). Solid lines represent quark propagators, dashed lines meson propagators. The crosses represent regulator insertions $\partial_kR_k$ and the red circles the appropriate vertex functions.}
	\label{fig:flow_Gamma2}
\end{figure*}

In order to obtain the flow equation for the quark two-point function we will follow the approach presented in Ref.\ \cite{TripoltWeyrichSmekalEtAl2018}, but extend it to finite spatial momentum, finite temperature, and quark chemical potential. First, we take two functional derivatives of the Wetterich equation, Eq.~(\ref{eq:Wetterich}), with respect to the fermionic fields, which yields
\begin{align}
\label{eq:flow_gamma2}
&\partial_k\Gamma_{k,\psi}^{(2),E}(p)=\frac{1}{2}\Tr\Big(\partial_k R_B(\vec{q}-\vec{p})D_{\phi\phi}(q-p)\nonumber\\
&\hspace*{2.5cm}\times \Gamma^{(3)}_{\bar\psi \psi \phi}D_{\bar\psi \psi}(q)\Gamma^{(3)}_{\bar\psi \psi \phi}D_{\phi \phi}(q-p)\nonumber\\
&\hspace*{2.5cm}+\partial_k R_F(\vec{q}+\vec{p})D_{\bar\psi \psi}(q+p) \nonumber \\
& \hspace*{2.5cm} \times \Gamma^{(3)}_{\bar\psi \psi \phi}D_{\phi\phi}(q)\Gamma^{(3)}_{\bar\psi \psi \phi}D_{\bar\psi \psi}(q+p)\Big),
\end{align}
see Fig.~\ref{fig:flow_Gamma2} for a diagrammatic representation. Therein, $q=(q_0,\vec{q})$ is the internal and $p=(p_0,\vec{p})$ the external momentum, $D=(\Gamma_{k}^{(2)}+R_k)^{-1}$ is the full regulated propagator, the vertex functions $\Gamma^{(3)}_{\bar\psi \psi \phi}$ are obtained from the ansatz in Eq.~(\ref{eq:gamma}), and the remaining trace represents a summation over all internal indices as well as an integration over the internal momentum, see App.~\ref{app:FRG} for explicit expressions. As in the original studies \cite{Kamikado2014,Tripolt2014,Tripolt2014a, TripoltWeyrichSmekalEtAl2018}, we use three-dimensional regulator functions, which only regulate spatial momenta but not the energy components, at the expense of slightly breaking the Euclidean $O(4)$ symmetry \cite{Kamikado2014}. While in principle also four-dimensional regulator functions can be used \cite{Pawlowski:2015mia,Pawlowski:2017gxj}, the three-dimensional regulators allow to analytically perform the integration over the internal energy component, or the corresponding Matsubara sum at finite temperature, which significantly simplifies the analytic-continuation procedure discussed in the following.

In order to obtain the flow equation for the real-time quark two-point function, we have to perform an analytic continuation from imaginary to real energies. This analytic continuation is performed on the level of the flow equations and is achieved by the following two-step procedure, see also Refs.\  \cite{Tripolt2014,Tripolt2014a}. First, the periodicity of the bosonic and fermionic occupation numbers, which appear in the flow equation upon evaluating the Matsubara summation analytically, with respect to the Euclidean Matsubara frequencies $p_0=i\,2\pi n T$ is exploited,
\begin{align}
n_{B,F}(E+i p_0)\rightarrow n_{B,F}(E).
\end{align}
In a second step, $p_0$ is replaced by a continuous real energy $\omega$,
\begin{align}
\label{eq:continuation2}
\partial_k\Gamma^{(2),R}_{k,\psi}(\omega,\vec p)=-\lim_{\epsilon\to 0} \partial_k\Gamma^{(2),E}_{k,\psi}(p_0=-i(\omega+i \epsilon), \vec p).
\end{align}
One should note that the limit $\epsilon\to 0$ can be taken analy\-tically for the imaginary part of the two-point function, while for the real part we use a small numerical value $\epsilon=1$~MeV. This analytic-continuation procedure obeys the physical Baym-Mermin boundary conditions \cite{Baym1961, Landsman1987}, and the resulting retarded propagator is analytic in the upper half of the complex-energy plane, as expected. 

We now make the following ansatz for the scale-dependent quark two-point function,
\begin{align}
\label{eq:Gamma2_R}
\Gamma^{(2)}_{k,\psi}(\omega,\vec{p})=\gamma_0C_k(\omega,\vec{p}) + i\vec{\gamma} \cdot \hat{p}\:A_k(\omega,\vec{p})- B_k(\omega,\vec{p})\, ,
\end{align}
with $\hat{p}\equiv\vec{p}/|\vec{p}|$ and where the UV initial conditions for the dressing functions are given by
\begin{align}
\label{eq_UV_values_coeffs}
A_{\Lambda}(\omega,\vec{p}) &= |\vec{p}|\,,\\
B_{\Lambda}(\omega,\vec{p}) &= h\sigma\, , \label{initialB} \\
C_{\Lambda}(\omega,\vec{p}) &= \omega + \mu \, , \label{initialC}
\end{align}
see also Ref.\ \cite{TripoltWeyrichSmekalEtAl2018}. The flow equation for the quark two-point function,
\begin{align}
\partial_k\Gamma^{(2)}_{k,\psi}(\omega,\vec{p})=&\gamma_0\partial_kC_k(\omega,\vec{p}) + i\vec{\gamma} \cdot \hat{\vec{p}}\:\partial_kA_k(\omega,\vec{p})\nonumber\\
&\quad -\partial_kB_k(\omega,\vec{p})\, ,
\end{align}
then leads to flow equations for the individual dressing functions, which are discussed in the appendix. In particular, the analyticity of the flow of these dressing functions in the upper half of the complex-energy plane is evident from these expressions, cf.~Eqs.~(\ref{eq:JA1})-(\ref{eq:JC2}), and guarantees that the correct analytic behavior of the retarded propagator is maintained in the flow.

\subsection{Quark spectral function}
\label{sec:spectral}

The quark spectral function is then given by
\begin{align}
\rho_{k,\psi}(\omega,\vec{p})=-\frac{1}{\pi}\Im G_{k,\psi}(\omega,\vec{p})\,,
\end{align}
where the propagator is defined as the inverse of the two-point function. The quark spectral function therefore has the same Dirac structure as the two-point function,
\begin{align}
\rho_{k,\psi}(\omega,\vec{p})=\gamma_0\rho^{(C)}_{k,\psi}(\omega,\vec{p})+i\vec{\gamma} \cdot \hat{p}\:\rho^{(A)}_{k,\psi}(\omega,\vec{p})+\rho^{(B)}_{k,\psi}(\omega,\vec{p})\, ,
\end{align}
where the individual components can be obtained as
\begin{align}
\rho^{(X)}_{k,\psi}(\omega,\vec{p})&=- \frac{1}{\pi}\Im G^{(X)}_{k,\psi}(\omega,\vec{p})\, ,
\end{align}
with $X\in \{A,B,C\}$. For a discussion of the properties of these spectral functions and the corresponding sum rules we refer to Ref.\ \cite{TripoltWeyrichSmekalEtAl2018}.

In the following, we will focus on particle and anti-particle spectral functions, i.e., states associated with positive and negative energies, which can be extracted from the quark spectral function $\rho_{k,\psi}(\omega,\vec{p})$ by applying suitable projection operators. A general form of such a projection operator is given by 
\begin{align}
\Lambda^{\pm}=\frac{1}{2\epsilon_p}\left[ \epsilon_p\pm (\gamma_0\vec{\gamma} \cdot \vec{p}+m\gamma_0)\right]\, ,
\end{align}
with $\epsilon_p=\sqrt{\vec{p}^{\,2}+m^2}$, see for example Refs.\ \cite{BlaizotOllitrault1993, Weldon2000}. This projection operator, however, depends on the quark mass, which introduces a certain ambiguity when dealing with resonance states as encountered in the following. We will therefore focus on two special cases where the dependence on the quark mass drops out: the case of zero momentum and that of zero quark mass. For $\vec{p}=0$, the quark spectral function can be decomposed as
\begin{align}
\rho_{k,\psi}(\omega,0)=\rho_{k,L}^+(\omega)L_+\gamma_0+ \rho_{k,L}^-(\omega)L_-\gamma_0,
\end{align}
with $L_\pm=(1\pm\gamma_0)/2$, while for $m_\psi=h\sigma_0=0$, we have
\begin{align}
\label{defrhop}
  \rho_{k,\psi}(\omega,\vec{p})=\rho_{k,P}^+(\omega,\vec{p})P_+\gamma_0+ \rho_{k,P}^-(\omega,\vec{p})P_-\gamma_0,
\end{align}
with $P_\pm=(1\pm\gamma_0\,\vec{\gamma}\cdot \hat{p})/2$. For further details and properties of these spectral functions we refer to App.~\ref{app:FRG}.

\section{FRG-improved one-loop calculation}
\label{sec:1_loop}

It will be instructive to compare the results obtained with the FRG setup to those from a one-loop calculation. In order to arrive at a meaningful comparison, we will use the same parameters and masses in the one-loop calculation as in the FRG calculation. This in particular entails making the meson masses in the one-loop calculation momentum-dependent, where the momentum scale is identified with the FRG scale, as discussed in the following, see also Ref.\ \cite{WangHe2018}.

We first write the retarded quark propagator as
\begin{align}
G_\psi^R(\omega,\vec{p})
&=\left[(\omega+i\epsilon+\mu)\gamma_0-m_\psi+i\vec{p}\,\vec{\gamma}-\Sigma^R(\omega,\vec{p})\right]^{-1},\nonumber \\
&=\left[\gamma_0C(\omega,\vec{p})+i\vec{\gamma}\hat{p}\:A(\omega,\vec{p})-B(\omega,\vec{p})\right]^{-1}\, ,
\end{align}
where $\Sigma^R(\omega,\vec{p})$ is the quark self-energy. The real-time quark self-energy can be obtained from its Euclidean counterpart $\Sigma^E(i\omega_n,\vec{p})$ by analytic continuation $i\omega_n\rightarrow \omega+i\epsilon$, where $\omega_n=(2n+1)\pi T$ are the fermionic Matsubara frequencies. In the imaginary-time formalism we can write the quark self-energy as
\begin{eqnarray}
\lefteqn{2\,\Sigma^E(i\omega_n, \vec{p})} \nonumber \\
& = & - h^2 T\sum_{m}\int \frac{d^3q}{(2\pi)^3}
\Big[ G_\sigma(i\omega_{m}-i\omega_{n},\vec{q}-\vec{p})G_\psi(i\omega_{m},\vec{q})\nonumber \\
& & \hspace*{1cm} +G_\sigma(i\nu_{m},\vec{q})G_\psi(i\nu_{m}+i\omega_{n},\vec{q}+\vec{p})\nonumber \\
& &\hspace*{1cm} +3G_\pi(i\omega_{m}-i\omega_{n},\vec{q}-\vec{p})i\gamma_5G_\psi(i\omega_{m},\vec{q})i\gamma_5\nonumber \\
& &\hspace*{1cm}  +3G_\pi(i\nu_{m},\vec{q})i\gamma_5G_\psi(i\nu_{m}+i\omega_{n},\vec{q}+\vec{p})i\gamma_5
\Big]\,, \nonumber \\
\label{eq:self_energy}
\end{eqnarray}
where $h$ is the Yukawa coupling and $\nu_{m}=2m\pi T$ are the bosonic Matsubara frequencies. We note that, in order  to facilitate a comparison, the momentum routing is chosen to be the same as in the FRG case, see also Fig.~\ref{fig:flow_Gamma2}. This also gives rise to the overall factor 2 in Eq.~(\ref{eq:self_energy}), since we are using four instead of the usual two diagrams. The right-hand side of Eq.~(\ref{eq:self_energy}) contains the free quark and meson propagators, which are given by
\begin{align}
G_\psi(i\omega_{n},\vec{p})&=\left[(i\omega_{n}+\mu)\gamma_0+i\vec{p}\cdot\vec{\gamma}-m_\psi\right]^{-1},\\
G_\alpha(i\nu_{n},\vec{p})&=\left[(i\nu_{n})^2-\vec{p}^{\,2}-m_\alpha^2(|\vec{p}|)\right]^{-1},
\end{align}
with $\alpha\in \{\sigma,\pi\}$ and where the meson masses are taken to be momentum-dependent. The momentum dependence is here identified with the scale dependence from the FRG calculation, i.e., $m_\alpha(|\vec{p}|)\equiv m_\alpha(k)$. The other parameters, such as the quark mass $m_\psi=h\sigma$, the Yukawa coupling $h$, the UV cutoff $\Lambda$, and the initial values for the propagator at the cutoff scale are also taken to be the same as in the FRG calculation, in order to facilitate a comparison. 

With this ``FRG-improved'' one-loop setup we can obtain the quark spectral function from the same expressions as discussed for the FRG case, but with the coefficients $X\in \{A,B,C\}$ replaced by $X(\omega,\vec{p})=X_\Lambda(\omega,\vec{p})+\Delta X(\omega,\vec{p})$ with
\begin{align}
\label{eq:one_loop_coeffs}
\Delta X(\omega,\vec{p})&=
\mathcal{L}^{(X)}_{\sigma\psi}(\omega,\vec{p})+
\mathcal{L}^{(X)}_{\psi\sigma}(\omega,\vec{p})\nonumber\\
&\quad \: \: \: +
3\,\mathcal{L}^{(X)}_{\pi\psi}(\omega,\vec{p})+
3\,\mathcal{L}^{(X)}_{\psi\pi}(\omega,\vec{p})\, ,
\end{align}
where explicit expressions for the loop functions $\mathcal{L}^{(X)}$ are given in App.~\ref{app:1_loop}. As shown in the appendix, there is a striking similarity between the one-loop equations for these loop functions and the corresponding FRG loop functions. In fact, the FRG equations can be obtained by taking a derivative of the one-loop equations, at least for zero momentum, where it can also be shown that the final result for the coefficients $X\in \{B,C\}$ only differs by a boundary term, see App.~\ref{app:relation} for a detailed discussion.

\section{Results}
\label{sec:results}

\subsection{Decay channels and quasi-particle energies}
\label{sec:processes}

\begin{figure}[t!]
	\includegraphics[width=\columnwidth]{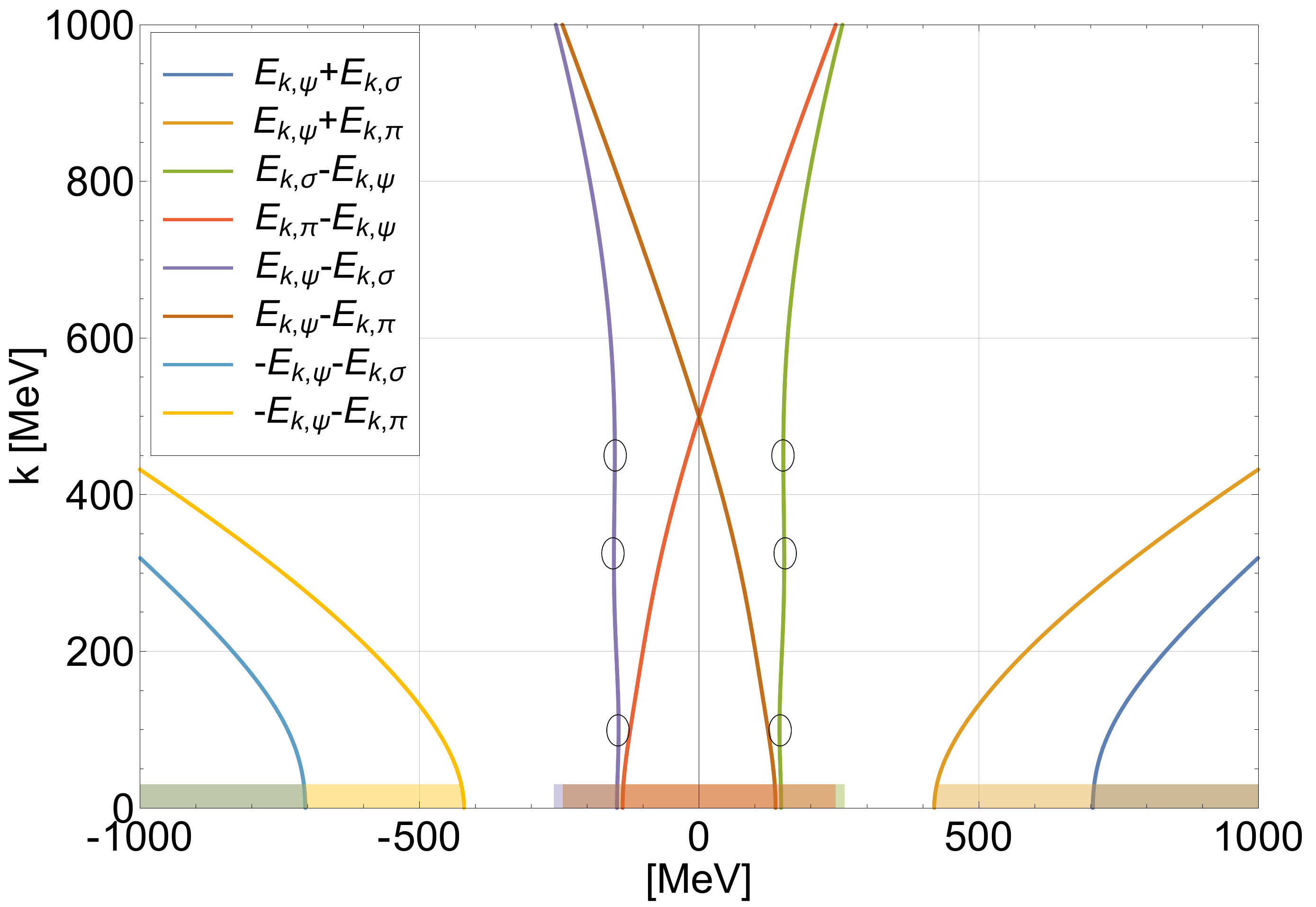}\\[-2mm]
	\caption{Combinations of the quasi-particle energies are shown vs.~the RG scale $k$ at $T=100$~MeV, $\mu=0$ and $|\vec{p}|=0$. Circles indicate the locations of van Hove-like points while continuum regions are indicated by  colored bars, see text for details.}
	\label{fig:energies_flow}
\end{figure}

\begin{figure*}[t!]
	\includegraphics[width=\columnwidth]{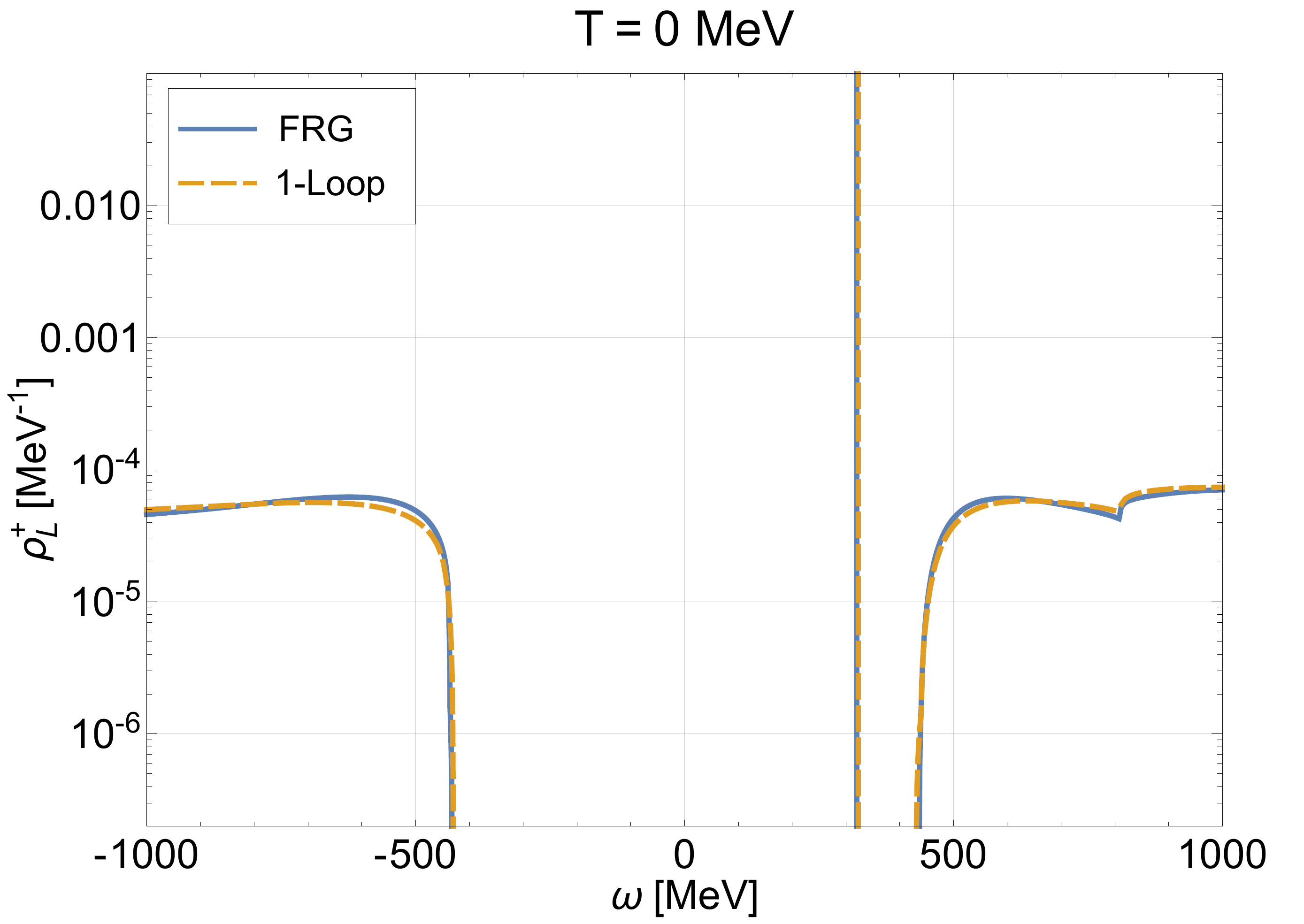}
	\includegraphics[width=\columnwidth]{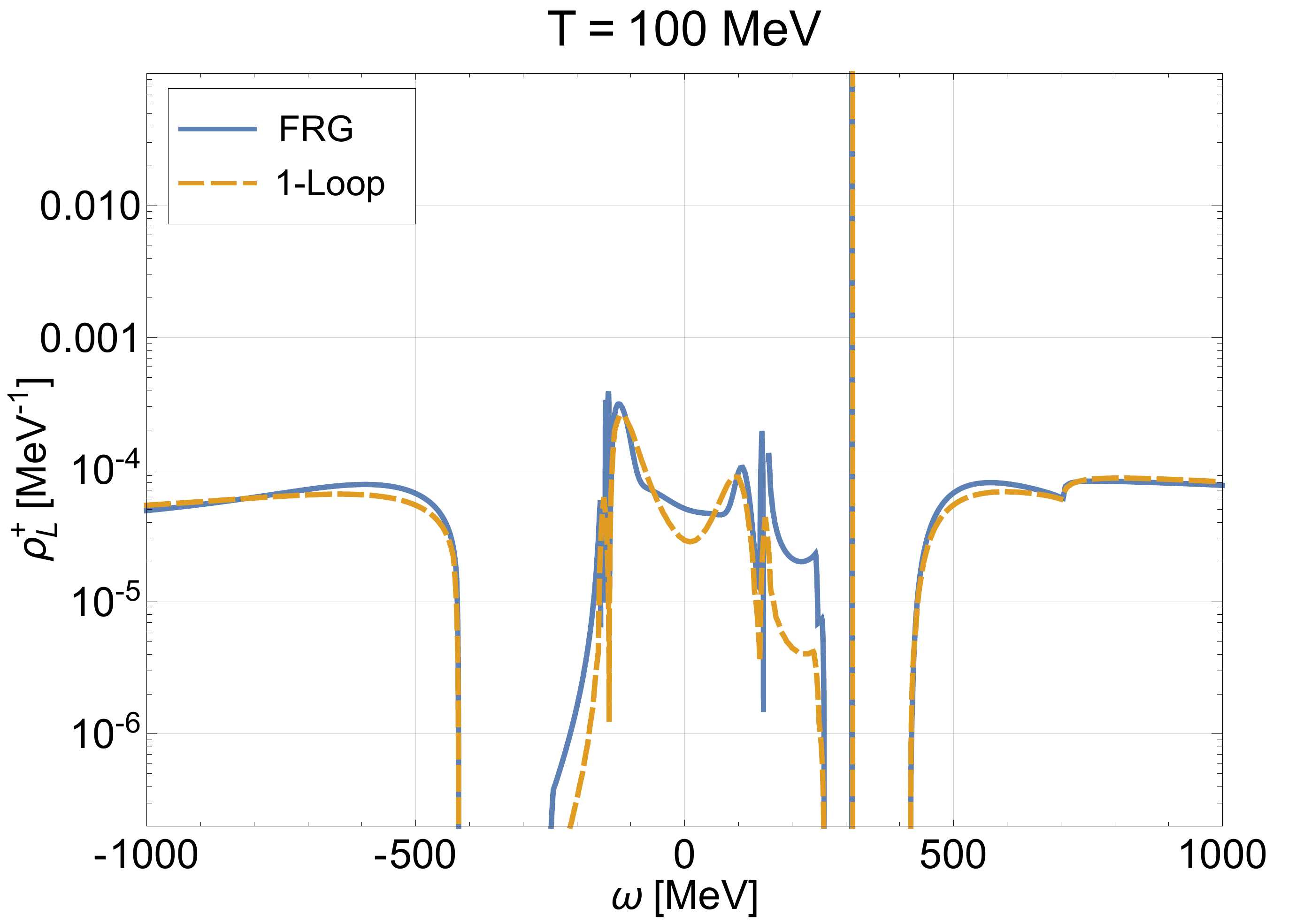}\\[2mm]
	\includegraphics[width=\columnwidth]{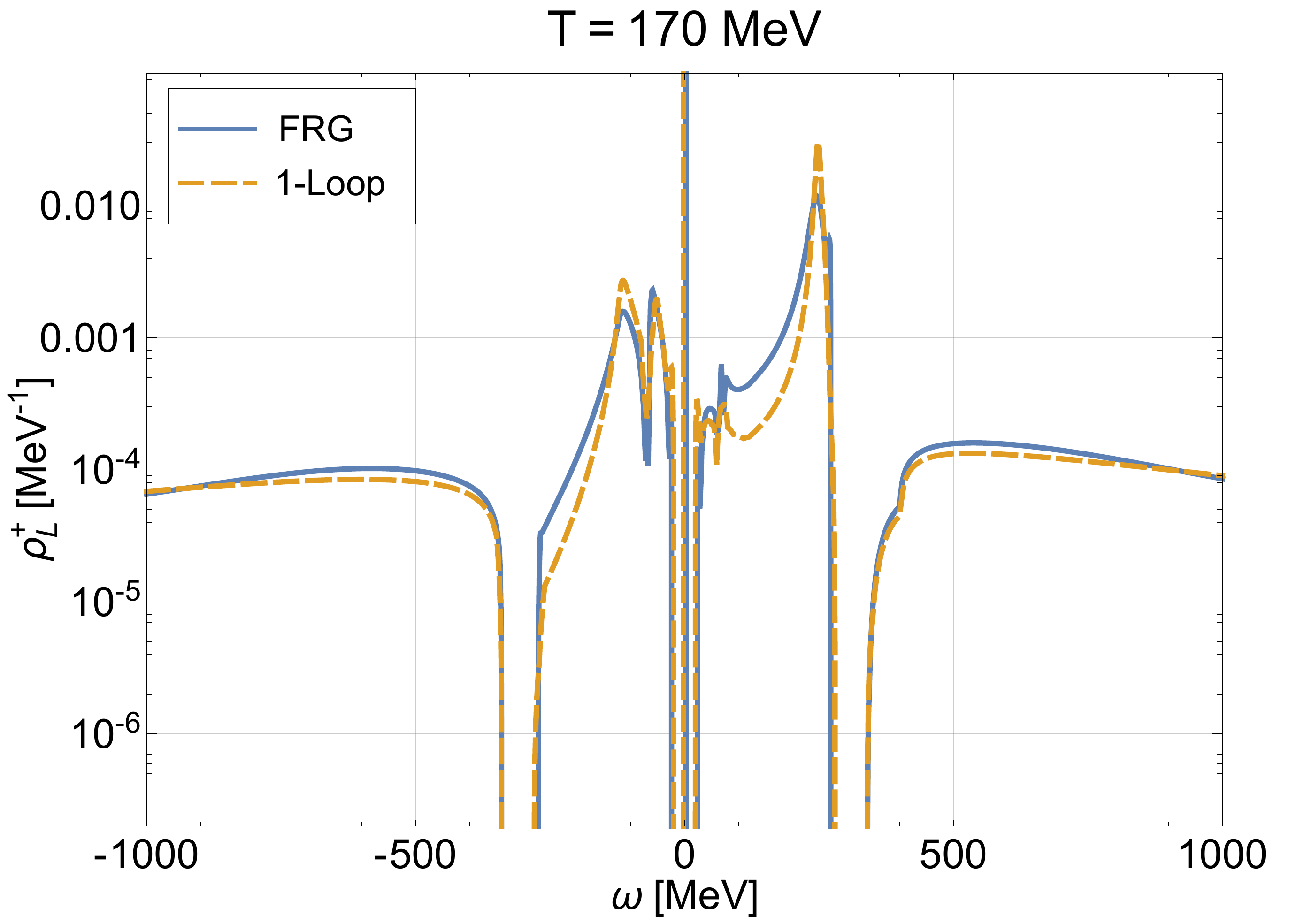}
	\includegraphics[width=\columnwidth]{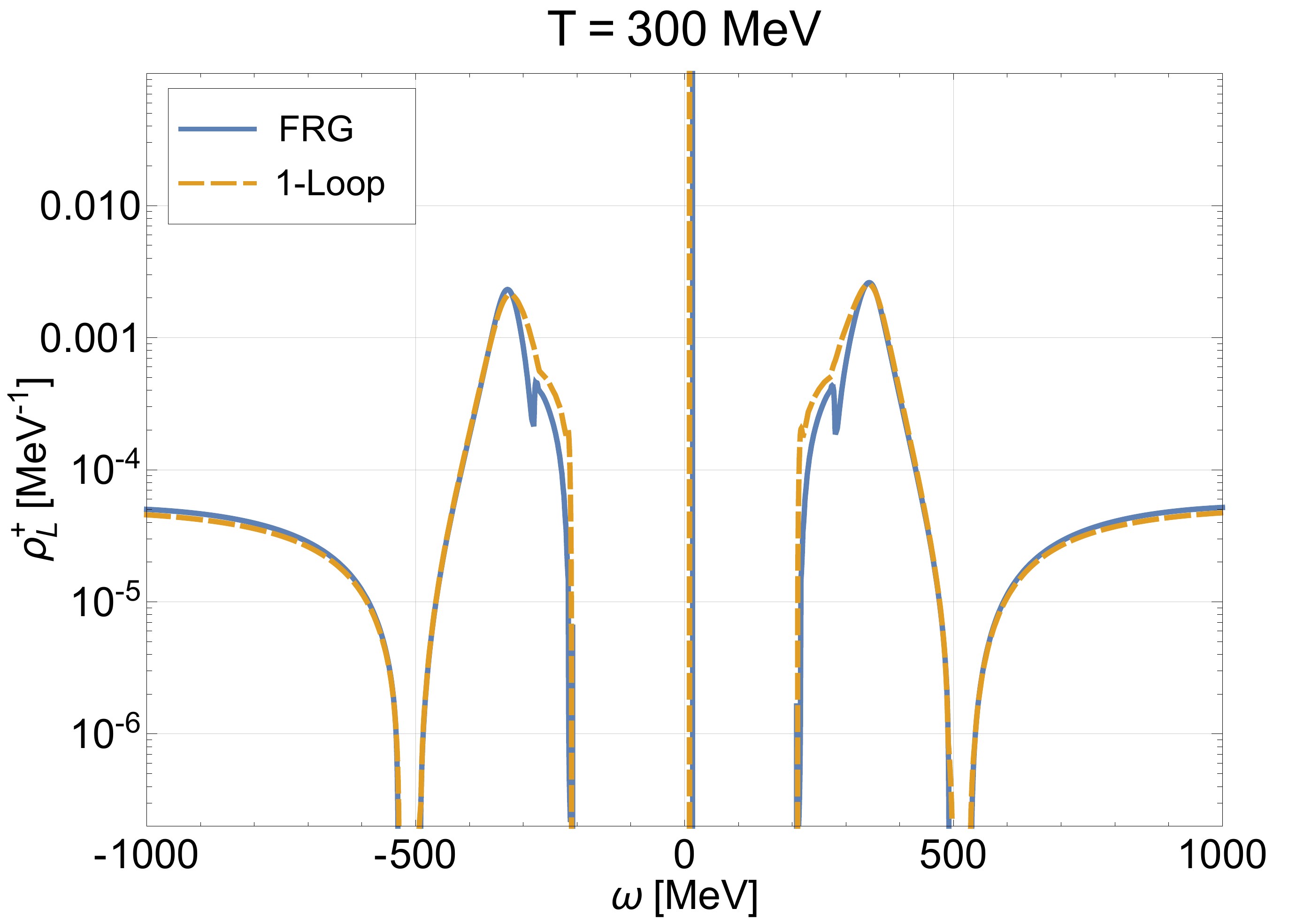}\\[-3mm]
	\caption{The quark spectral function $\rho^+_L(\omega)$ as obtained from the FRG (solid) in comparison to the FRG-improved one-loop result (dashed) at $\mu=0$ and different temperatures: $T=0, 100, 170$, and $300$~MeV, see text for details.}
	\label{fig:spectral_T}
\end{figure*}

The structure of the quark spectral functions as obtained from the FRG (or the FRG-improved one-loop setup) depends to a large part on the behavior of the scale-dependent quasi-particle energies, since they determine scale-dependent thresholds $\omega^c_k$ for continuum contributions to imaginary parts and hence the spectral functions, which are accumulated during the flow.  
In Fig.~\ref{fig:energies_flow} we show, as an illustrative example, the scale dependence of certain combinations of these energies at a temperature of $T=100$~MeV. The energies themselves are obtained by solving the flow equation for the effective potential, Eq.~(\ref{eq:flow_pot}), and monitoring the scale-dependent energies, as defined in Eqs.~(\ref{eq:energies})-(\ref{eq:masses}), at the IR minimum $\sigma_0$.

In particular, we show the following combinations of energies, connected with scale-dependent continuum thresholds $\omega^c_k$, as summarized here for $|\vec{p}|=0$:
\begin{align}
&\psi^* \leftrightarrow\psi+\alpha, &&  \quad \omega^c_k = E_{k,\psi}+E_{k,\alpha},\\
&\psi^*+\bar{\psi} \leftrightarrow\alpha, &&  \quad \omega^c_k = E_{k,\alpha}-E_{k,\psi},\\
&\psi^*+\alpha \leftrightarrow\psi, &&  \quad \omega^c_k = E_{k,\psi}-E_{k,\alpha},\\
&\psi^*+\bar{\psi} + \alpha \leftrightarrow 0, &&  \quad \omega^c_k = -E_{k,\psi}-E_{k,\alpha},
\end{align}
with $\alpha\in\{\sigma,\pi\}$. The first type of process, $\psi^* \leftrightarrow\psi+\alpha$, describes the decay of an off-shell quark state $\psi^*$ with energy $\omega$ into an on-shell quark and meson with energies $E_{k,\psi}$ and $E_{k,\alpha}$, as well as its inverse process. The second and third line describe thermal processes that can only occur in a medium, while the other two processes can also occur in the vacuum. The continuum regions connected to these processes are indicated by colored bars in Fig.~\ref{fig:energies_flow}. We note that the thermal processes can also give rise to van Hove-like singularities in the spectral functions, which occur when the derivatives of the energy differences with respect to the scale $k$ vanish, see also Ref.\ \cite{JungRenneckeTripoltEtAl2017}. This effect can in principle occur several times during the flow, see also Fig.~\ref{fig:energies_flow} where in fact several such van Hove-like singularities can be identified at different scales $k$. We also note that the UV and IR values of the energy combinations discussed above determine the location of thresholds in the spectral functions, as discussed below.

\subsection{Quark spectral function at finite temperature}
\label{sec:finiteT}

We now turn to the quark spectral function at finite temperature but zero chemical potential and zero spatial momentum. This case was also studied in Ref.\ \cite{WangHe2018} using the same setup as presented here. In Fig.~\ref{fig:spectral_T} we show the quark spectral function $\rho^+_L(\omega)$ as obtained from the FRG as well as the FRG-improved one-loop setup at $T=0$, 100, 170, and 300 MeV.\footnote{We note that at high temperatures, $T>170$~MeV, the FRG results sometimes show additional delta peaks with a very small spectral weight, i.e., at least ten times smaller compared to the main delta-peak contribution, which are treated as truncation artifacts by comparison with the one-loop results and are not shown in the figures.} We note that the underlying quark-meson model exhibits a chiral crossover transition at zero chemical potential and finite temperature. This crossover temperature was found to be  $T_c\approx 170$~MeV for the parameters and the setup chosen in this work, see also Ref.\ \cite{Tripolt2014}.

We observe an overall good agreement between both frameworks, with quantitative differences arising mostly in the thermal continuum regime at smaller energies. This thermal regime exhibits a complicated structure which is largely due to the effects discussed above, i.e., van Hove-like peaks as wells as IR and UV thresholds. In the vacuum, the quark pole mass is found to be $m_\psi^p\approx 320$~MeV and then slightly decreases, being $m_\psi^p\approx 300$~MeV at $T=100$~MeV. With increasing temperature we also observe that the thresholds associated to the processes $\psi^* \leftrightarrow\psi+\sigma$ and $\psi^* \leftrightarrow\psi+\pi$ approach each other. This is expected due to the progressing restoration of the spontaneously broken chiral symmetry which entails that the masses of the chiral partners, i.e., of the sigma meson and the pion, become degenerate at high temperatures, see also Ref.\ \cite{Tripolt2014}.

\begin{figure*}[t!]
	\includegraphics[width=\columnwidth]{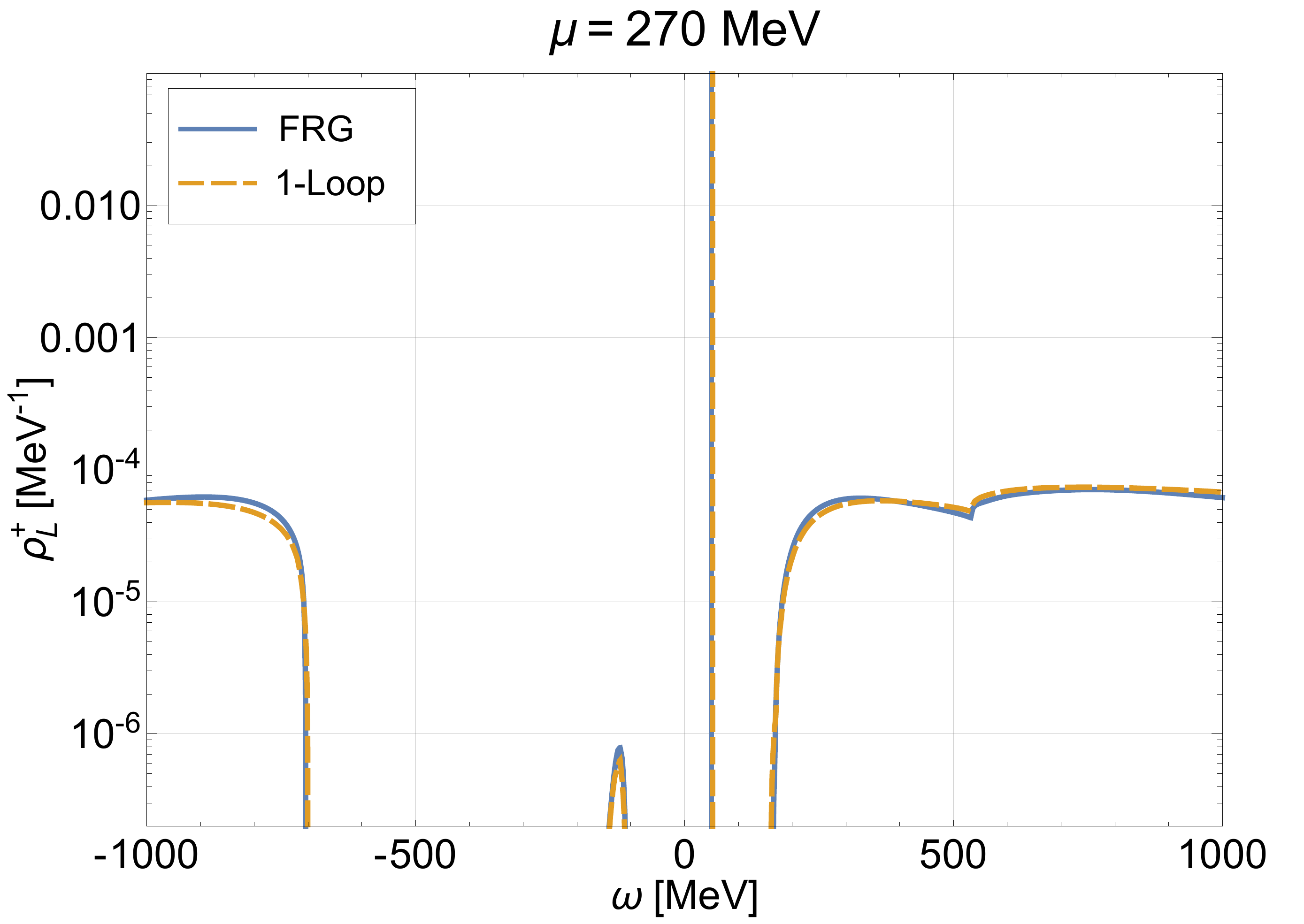}
	\includegraphics[width=\columnwidth]{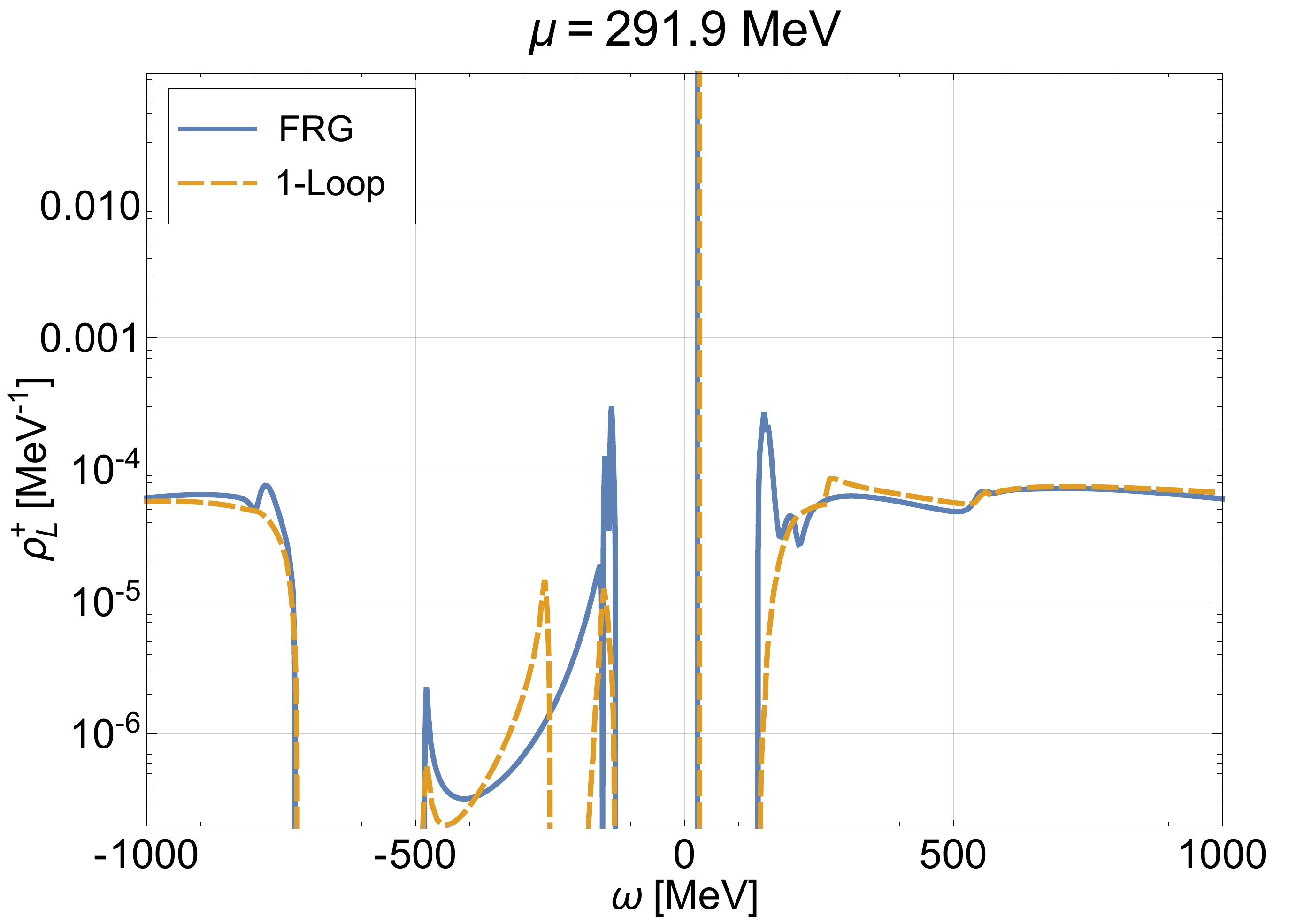}\\[-3mm]
	\caption{The quark spectral function $\rho^+_L(\omega)$ as obtained from the FRG (solid) in comparison to the FRG-improved one-loop calculation (dashed) at $T=9$~MeV and different chemical potentials: $\mu=270$~MeV (left) and $\mu=291.9$~MeV (right), see text for details.}
	\label{fig:spectral_mu}
\end{figure*}

\begin{figure}[t!]
	\includegraphics[width=\columnwidth]{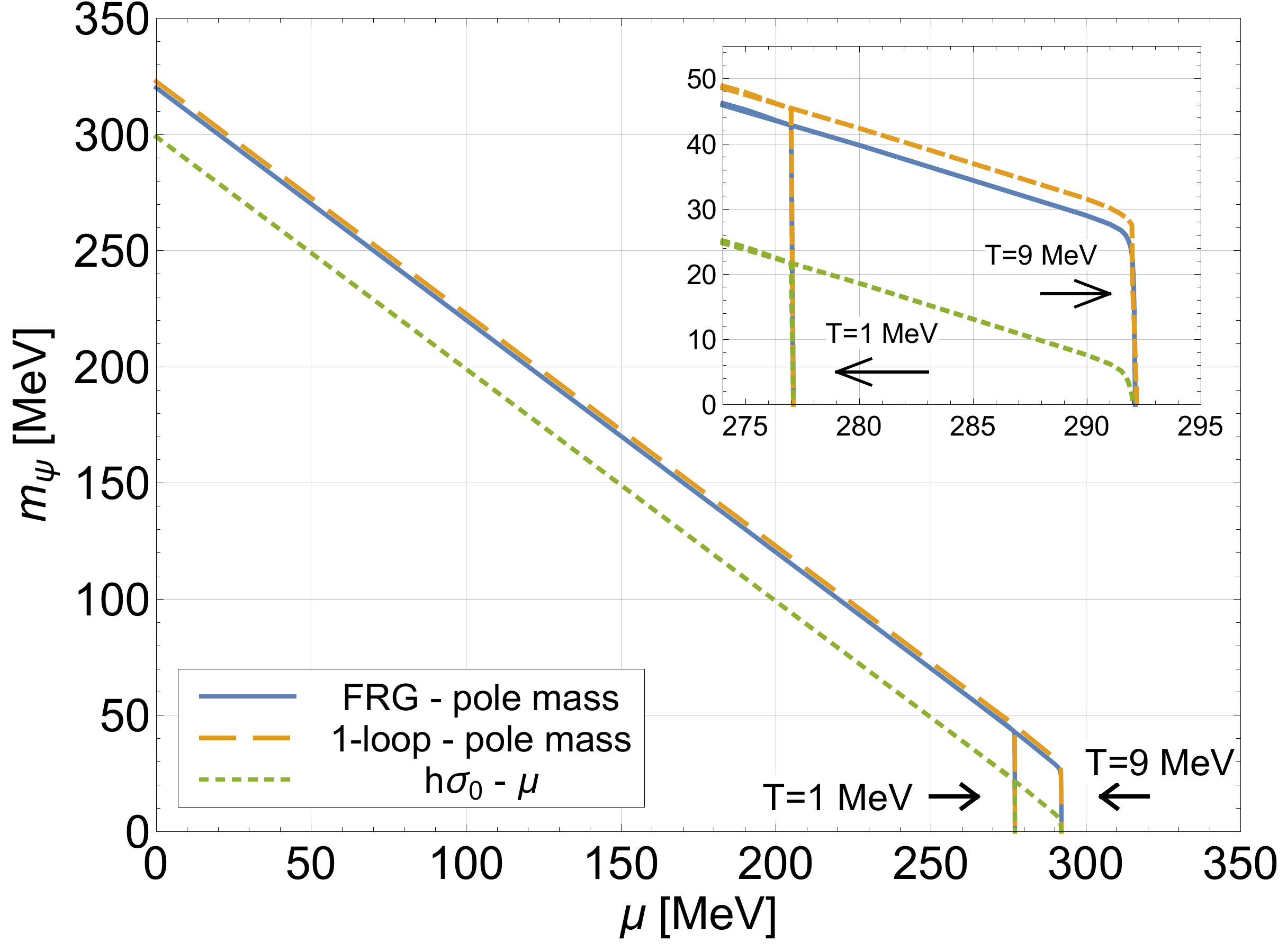}\\[-2mm]
	\caption{The quark pole mass $m_\psi^p$ as obtained from the FRG (solid) in comparison to the FRG-improved one-loop calculation (dashed), and the quark mass obtained from the minimum of the effective potential, $m_\psi=h\sigma_0$, (dotted) at $T=9$~MeV vs. chemical potential. In addition, we show the corresponding results for $T=1$~MeV where the masses exhibit a discontinuity at the first-order phase transition. The value of the quark mass at the discontinuity can be interpreted as the binding energy.}
	\label{fig:pole_mass}
\end{figure}

\begin{figure}[t!]
	\includegraphics[width=\columnwidth]{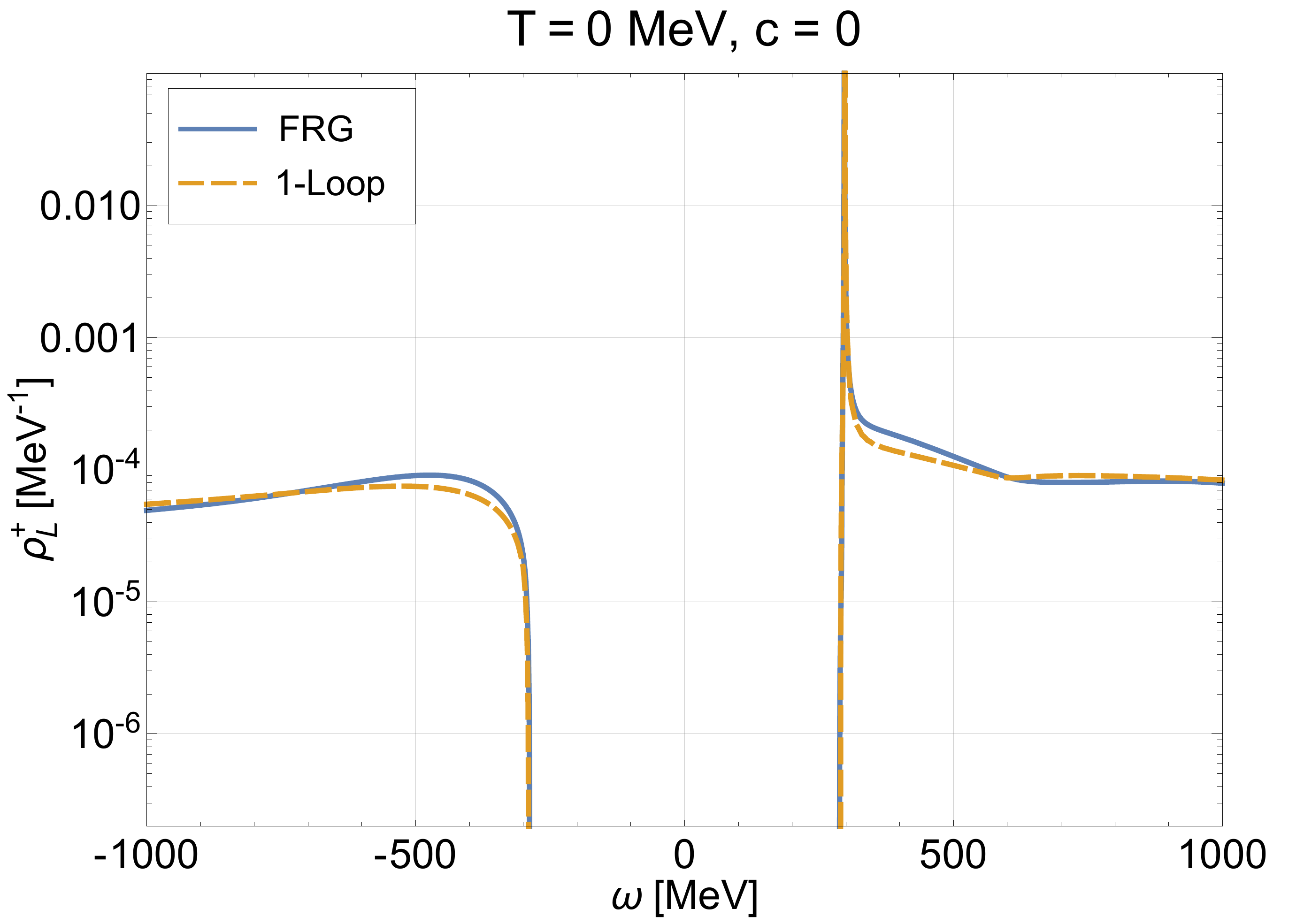}\\	\includegraphics[width=\columnwidth]{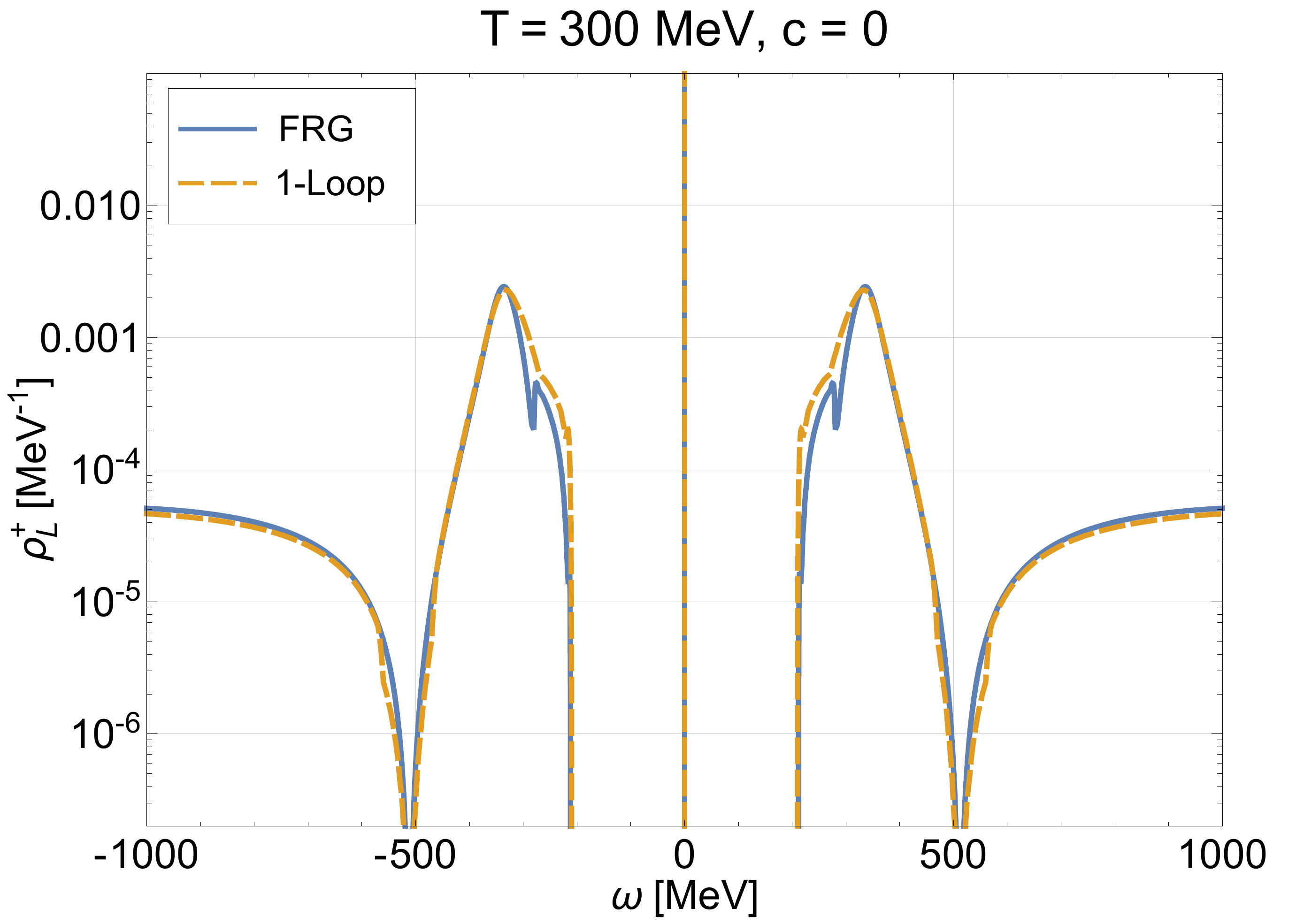}
	\caption{The quark spectral function $\rho^+_L(\omega)$ as obtained from the FRG (solid) as well as from the FRG-improved one-loop calculation (dashed) is shown at $T=0$~MeV (left) and $T=300$~MeV (right) for $\mu=0$ and $c=0$, i.e., in the chiral limit, see text for details.}
	\label{fig:spectral_c0}
\end{figure}

\begin{figure*}[t!]
	\includegraphics[width=1.6\columnwidth]{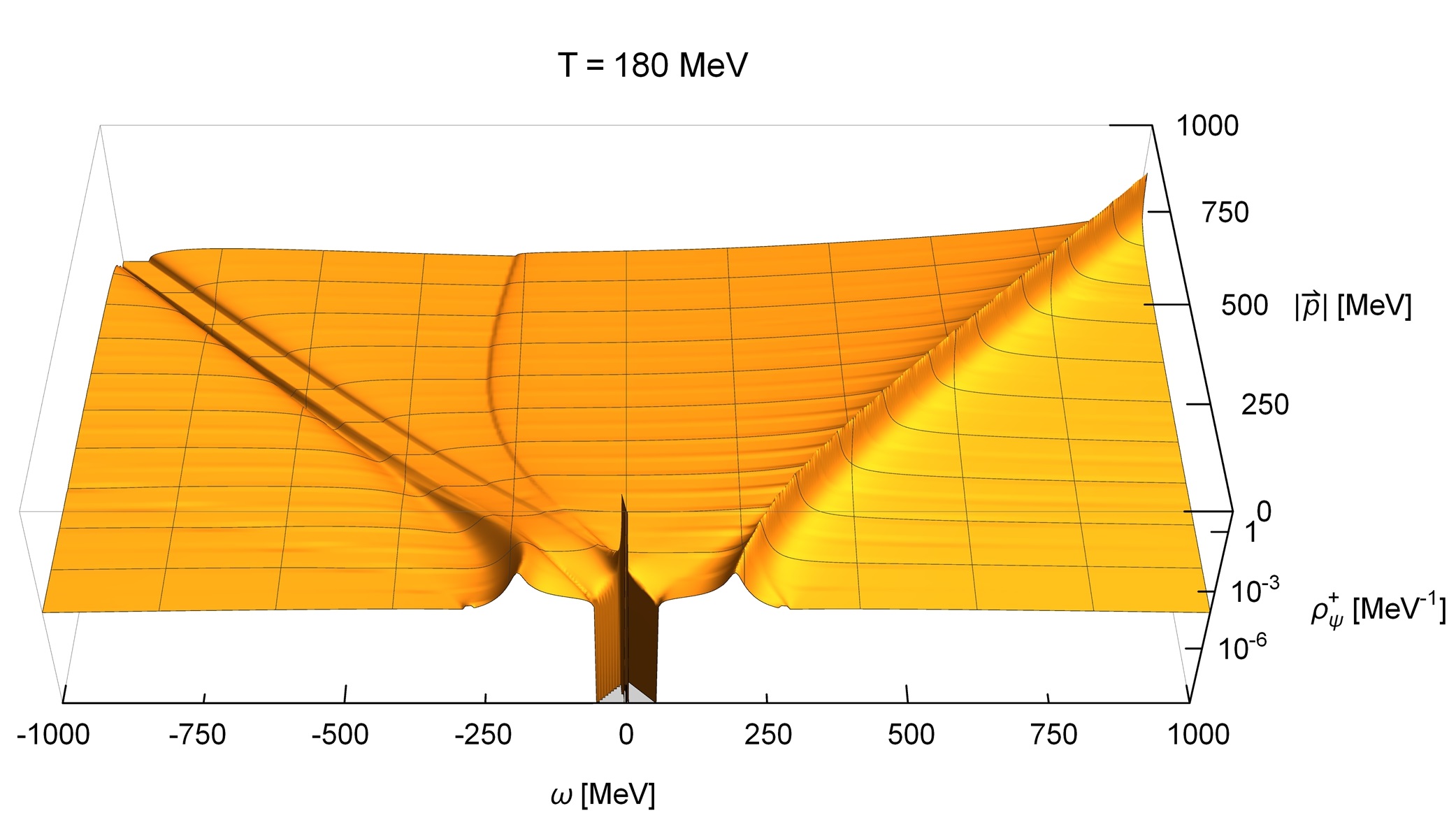}\\[-6mm]
	\includegraphics[width=1.6\columnwidth]{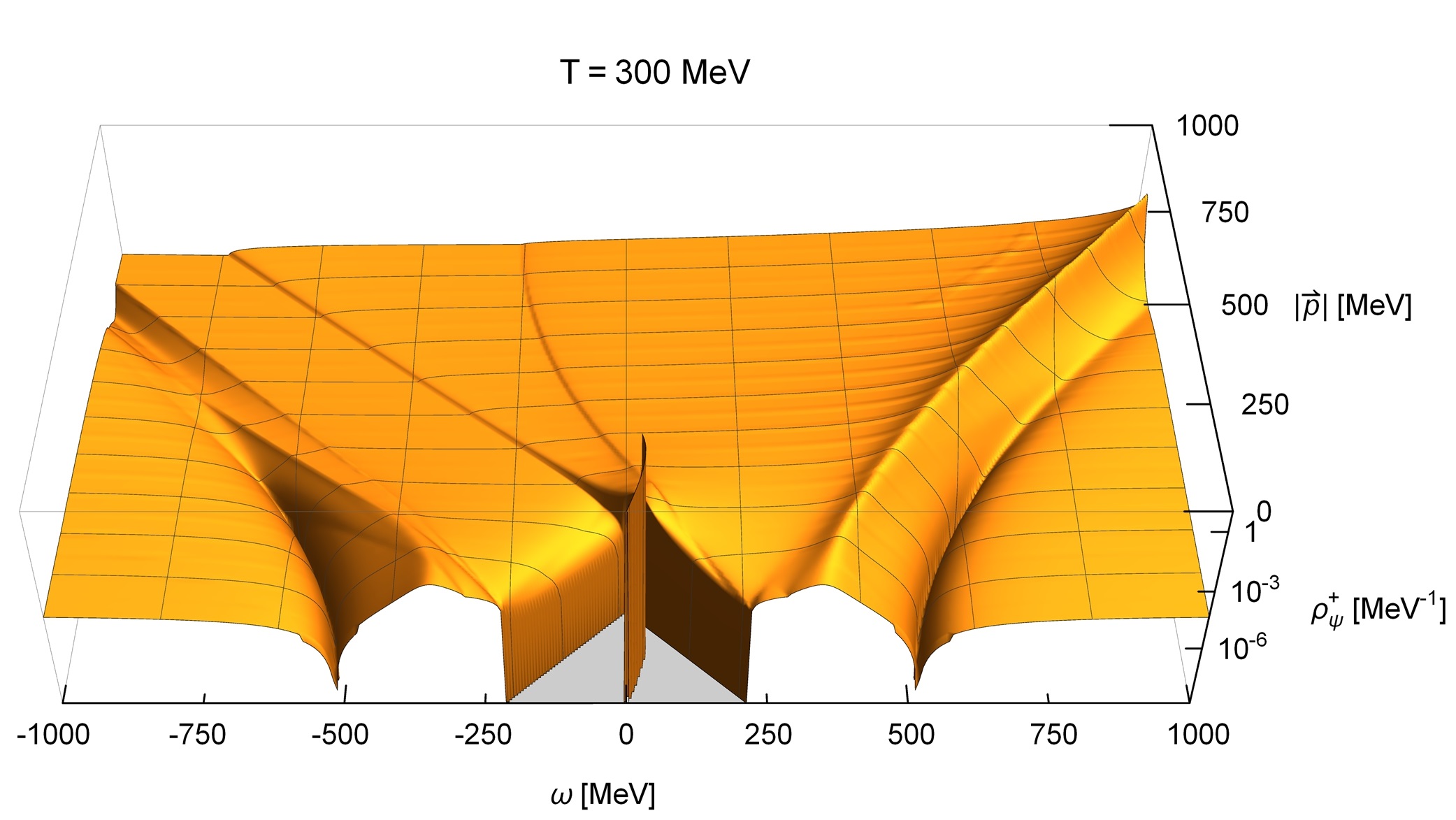}
	\caption{The quark spectral function $\rho^+_P(\omega,|\vec{p}|)$ as obtained from the FRG-improved one-loop calculation (dashed) is shown vs.~the external energy $\omega$ and the external spatial momentum $|\vec{p}|$ at $T=180$~MeV (top) and $T=300$~MeV (bottom) in the chiral limit, i.e., $c=0$, and for $\mu=0$, see text for details.}
	\label{fig:spectral_3D}
\end{figure*}

At $T\approx 170$~MeV we observe that the delta peak associated with the quark quasi-particle enters the thermal continuum regime, giving rise to a prominent peak with a finite width. At this temperature, i.e., at the chiral crossover temperature $T_c$ of the model, we also observe the formation of a massless mode at $\omega=0$~MeV. This so-called ``phonino'' mode acquires a small mass at $T>T_c$, with $m_{\text{phonino}}\approx 15$~MeV at $T=300$~MeV, while the thermal quasi-particle peak inside the continuum monotonically moves to higher energies with increasing temperature, being $m_\psi^p\approx 345$~MeV at $T=300$~MeV, cf.~Fig.~\ref{fig:spectral_T}. We find that the spectral weight of the phonino mode is about $50\%$ of the total weight of the spectral function, see also Ref.\ \cite{TripoltWeyrichSmekalEtAl2018}.

\subsection{Quark spectral function at finite density}
\label{sec:finitemu}

We now turn to the case of finite temperature and finite quark chemical potential but, for now, still zero spatial momentum. In Fig.~\ref{fig:spectral_mu} we show the quark spectral function $\rho^+_L(\omega)$ as obtained from the FRG as well as the FRG-improved one-loop setup at $T=9$~MeV and different chemical potentials. This temperature corresponds to the critical endpoint of the model, which was determined to be located at $T_{\text{CEP}}\approx9$~MeV and $\mu_{\text{CEP}}\approx 292$~MeV for the parameters chosen here \cite{TripoltSmekalWambach2017}. 

The left panel of Fig.~\ref{fig:spectral_mu} shows the quark spectral function at $\mu=270$~MeV. When comparing this with the vacuum spectral function in Fig.~\ref{fig:spectral_T}, we note that the chemical potential simply acts as an overall shift in the energy. This shift originates from the UV initial conditions for the dressing functions, cf.~Eq.~(\ref{initialC}), and from the structure of the flow equations, as presented in the appendix, where the energy $\omega$ always appears in combination with the chemical potential. The energy $\omega$ is therefore measured relative to the chemical potential and is interpreted as the additional energy needed to create a quark-like excitation, e.g.~given by the location of the delta peak in the spectral function. Apart from this shift, the quark spectral function remains almost unchanged over this wide range of chemical potentials because the temperature is so low, demonstrating that in our calculations there is no substantial Silver-Blaze problem \cite{Cohen2003}.

When approaching the CEP, however, non-trivial medium modifications due to the chemical potential have to become relevant, as shown in the right panel of Fig.~\ref{fig:spectral_mu}, i.e., at $\mu=291.9$~MeV. Apart from the thermal-continuum structure, which is now shifted to negative energies and arises due to the high quark density rather than the temperature, we observe that location and structure of the decay thresholds change when approaching the CEP. This particularly concerns the threshold connected with the process $\psi^* \leftrightarrow\psi+\sigma$, since the sigma mass rapidly decreases near the CEP and in fact vanishes at this second-order phase transition. The sigma-quark threshold therefore moves to smaller energies close to the CEP, cf.~Fig.~\ref{fig:spectral_mu}. We note that the study of the quark spectral function at chemical potentials beyond the CEP is deferred to future work since this region is hampered by the appearance of negative-entropy regimes within current FRG calculations, see also Ref.~\cite{Tripolt:2017zgc}.

We also find that close to the CEP, the quark pole mass decreases rapidly, see Fig.~\ref{fig:pole_mass}, where we show the dependence of the quark pole mass on the chemical potential for the FRG and the one-loop case. A deviation from the linear behavior only occurs very close to the CEP, where the quark pole mass as well as the Euclidean quark mass parameter obtained from the minimum of the effective potential (here plotted relative to the chemical potential) both rapidly drop and eventually vanish at the CEP. For comparison, we also show the corresponding results at $T=1$~MeV in Fig.~\ref{fig:pole_mass}. At this low temperature, the phase diagram exhibits a first-order phase transition, where the quark mass changes discontinuously, while at $T=9$~MeV the masses are continuous, see e.g.~Ref.\ \cite{Tripolt2014}. The height of the discontinuity determines the binding energy per quark of the self-bound quark matter that arises at this transition. 

\subsection{Quark spectral function at finite temperature and finite momentum in the chiral limit}
\label{sec:chiral_limit}

In the following we will study the quark spectral function in the chiral limit of vanishing explicit quark masses, i.e., at $c=0$. In this limit, the chiral crossover turns into a second-order phase transition where chiral symmetry becomes fully restored and the constituent quark mass vanishes continuously, cf.~e.g.~Ref.\ \cite{Schaefer:2004en}. On the other hand, in the chirally broken phase, i.e., for $T<T_c$ at $\mu=0$, the constituent quark mass is finite but the pions are massless. 

This can also be seen from Fig.~\ref{fig:spectral_c0} where we show the quark spectral function $\rho^+_L(\omega)$ as obtained from the FRG as well as the FRG-improved one-loop setup at  $T=0$~MeV (left) and $T=300$~MeV (right) for $\mu=0$ and $c=0$. We observe that, at $T=0$~MeV, the quark delta peak is now directly attached to the continuum generated by the process $\psi^* \leftrightarrow\psi+\pi$, since the pions are massless in this regime. At $T=300$~MeV, the quark spectral function is almost identical to the case with explicitly broken chiral symmetry, $c>0$, cf.~Fig.~\ref{fig:spectral_T}, and the quark quasi-particle peak has again merged with the thermal continuum. On the other hand, the phonino remains exactly massless for $T>T_c$ in the chiral limit.

\begin{figure*}[t!]
	\includegraphics[width=\columnwidth]{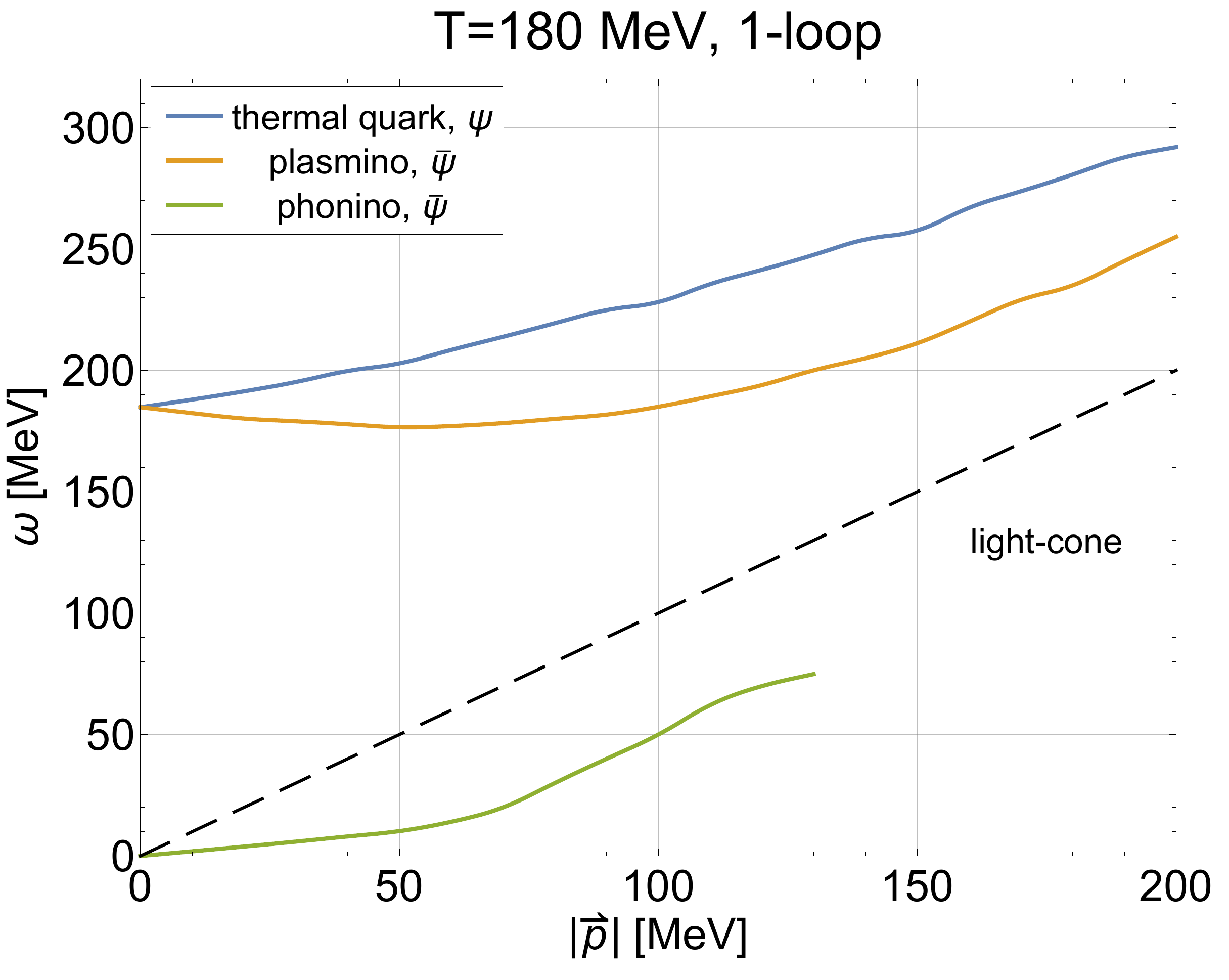}
	\includegraphics[width=\columnwidth]{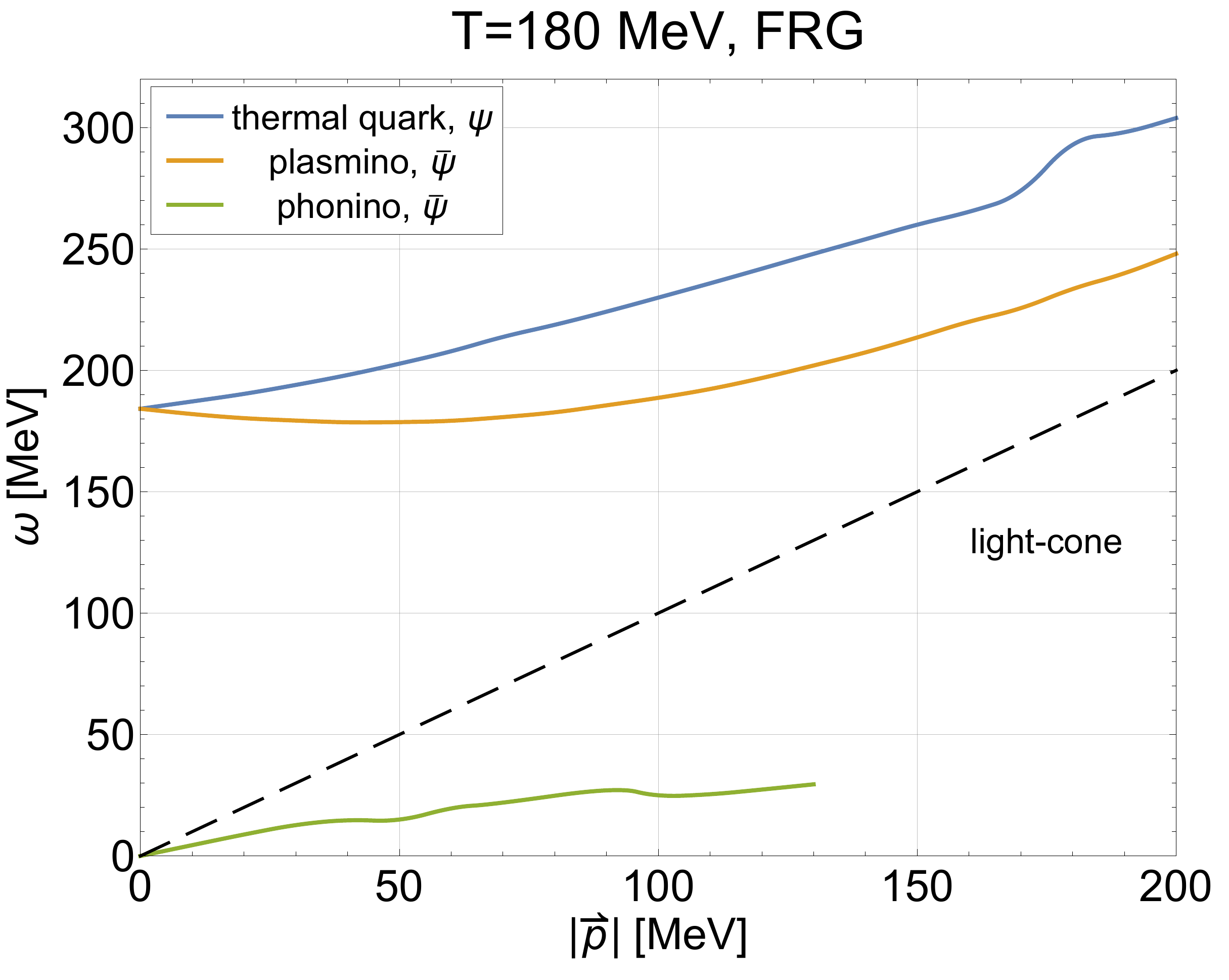}\\[2mm]
	\includegraphics[width=\columnwidth]{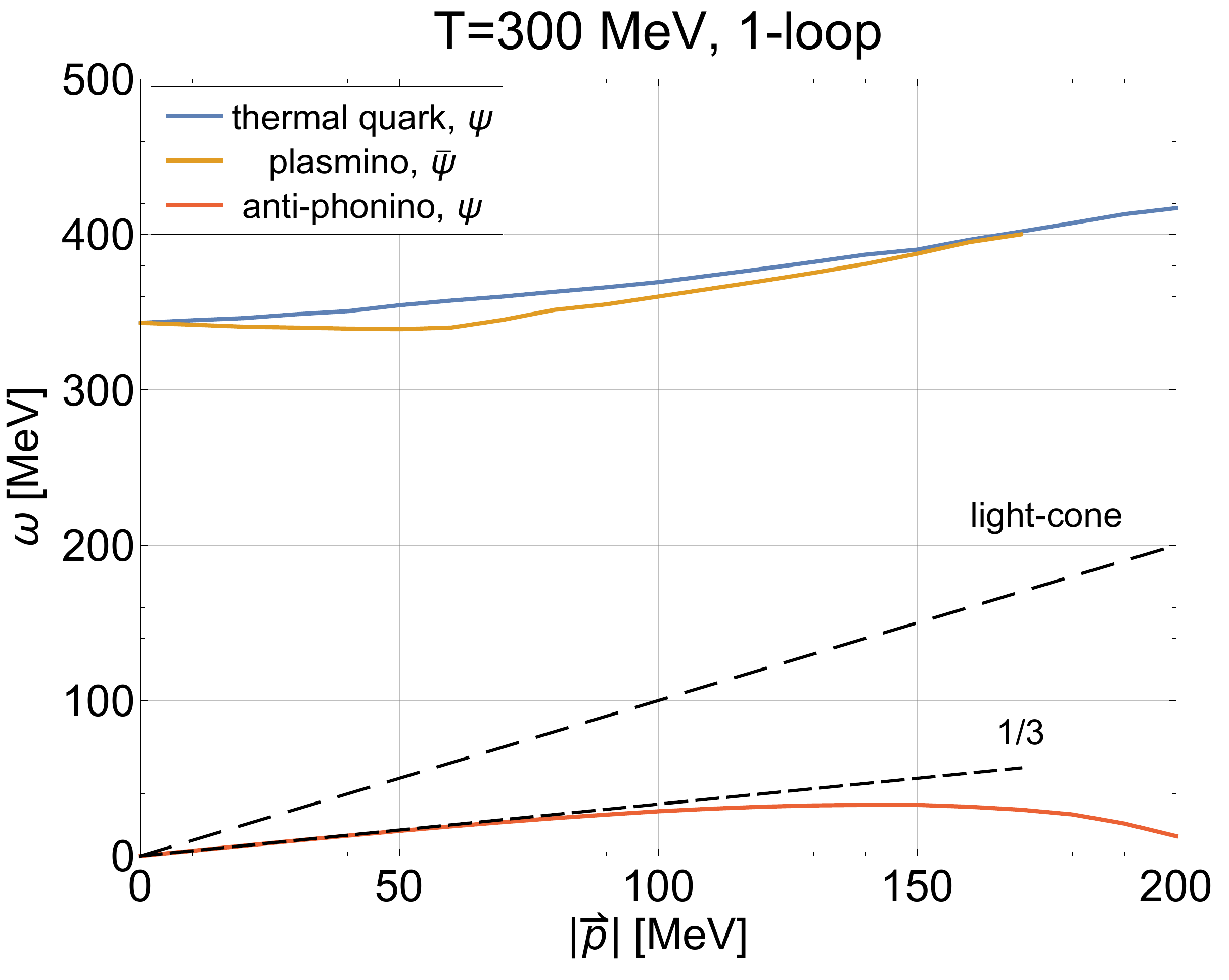}
	\includegraphics[width=\columnwidth]{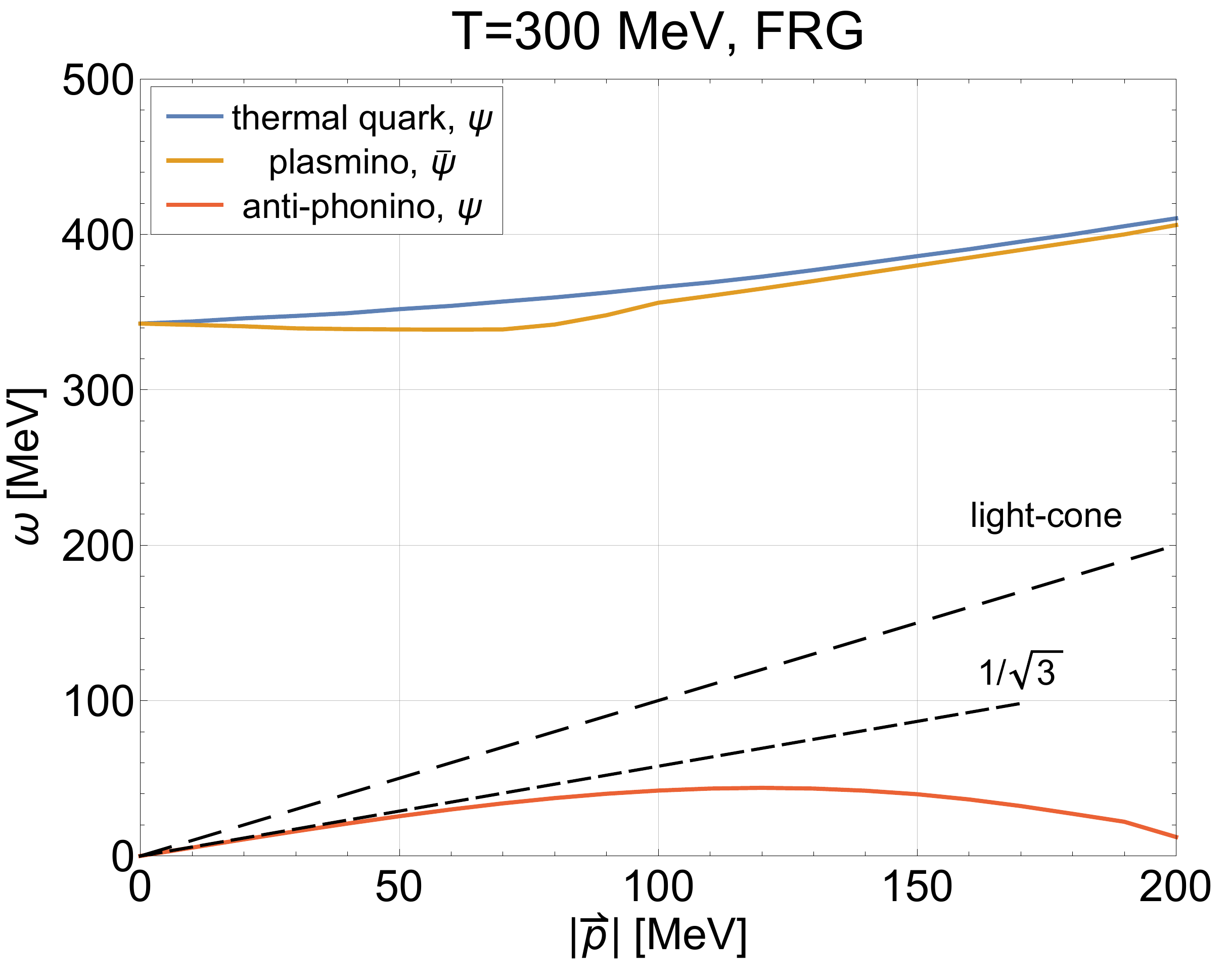}
	\caption{The dispersion relations of the identified fermionic excitations as obtained from the FRG-improved 1-loop calculation (left) and the FRG calculation (right) are shown at $T=180$~MeV (top) and at $T=300$~MeV (bottom) in the chiral limit, i.e., $c=0$, at $\mu=0$, see text for details.}
	\label{fig:dispersion}
\end{figure*}

We now turn to the case of finite external spatial momentum, $|\vec{p}|>0$, in the chiral limit at high temperatures. The particular advantage of this is that we can then use the definition of $\rho^+_P(\omega,|\vec{p}|)$ from Eq.~(\ref{defrhop}), since the quark mass is zero in the chirally restored phase at $T>T_c$. In Fig.~\ref{fig:spectral_3D}, we show the quark spectral function $\rho^+_P(\omega,|\vec{p}|)$ vs.~the external energy $\omega$ and the external spatial momentum $|\vec{p}|$ at $T=180$~MeV and $T=300$~MeV as obtained from the FRG-improved one-loop calculation. As a general observation we find that the particle peaks and thresholds are Lorentz-boosted to higher energies as the spatial momentum increases, as expected. Apart from the thermal quasi-particle peak at positive energies $\omega$ we can now also clearly identify the anti-plasmino mode at negative energies. At $|\vec{p}|=0$ these two modes are degenerate and located at the same value of $|\omega|$. We note that the actual plasmino mode does not show up in Fig.~\ref{fig:spectral_3D} since we project on particle-like excitations by using $\rho^+_P(\omega,|\vec{p}|)$. The plasmino, as well as the thermal anti-quark, however, carry the quantum numbers of an anti-particle. These modes can be obtained from $\rho^+_P(\omega,|\vec{p}|)$ by reflecting it at $\omega=0$, cf.~Eq.~(\ref{eq:rel2}).

We also observe the formation of the phonino mode at $T>T_c$. In fact, the delta peak at $\omega\approx 0$~MeV in Fig.~\ref{fig:spectral_3D} should rather be referred to as the anti-phonino mode, since the term phonino usually refers to a collective anti-particle-like excitation while Fig.~\ref{fig:spectral_3D} only shows particle-like excitations. The phonino itself can again be obtained by reflecting $\rho^+_P(\omega,|\vec{p}|)$ at $\omega=0$. We observe that at $T=180$~MeV the anti-phonino is boosted towards negative energies with increasing spatial momentum, as expected from other studies, see e.g.~Ref.\ \cite{BlaizotSatow2014}, while at $T=300$~MeV it is moving towards positive energies, until it merges with the thermal continuum in both cases.

\subsection{Fermionic excitations and dispersion relations}
\label{sec:dispersion}

We now turn to the dispersion relations of the fermionic excitations identified in the preceding section. We note that we use the peak positions in the spectral functions to define the corresponding dispersion relations $\omega(|\vec{p}|)$. In Fig.~\ref{fig:dispersion} we show the dispersion relations of the thermal quark, the plasmino, the phonino, and the anti-phonino at high temperatures in the chiral limit. The particle or anti-particle-like character of these excitations is indicated as $\psi$ or $\bar{\psi}$ in the figure.

At $T=180$~MeV we find that the dispersion relation of the thermal quark monotonically increases while the energy of the plasmino decreases at first and reaches a minimum at $|\vec{p}|\approx 50$~MeV, from where on it increases again and eventually approaches the thermal quark branch from below. This behavior is expected from standard HTL calculations where the slopes of these two branches are obtained from
\begin{align}
\omega=M \pm \frac{|\vec{p}|}{3}+\frac{|\vec{p}|^2}{3M}+\dots,
\end{align}
where $M$ is the mass of the thermal quark, see for example Ref.\ \cite{Lebellac:1996}. The phonino mode also shows an approximately linear dispersion relation at small momenta, until it merges with the thermal continuum at $|\vec{p}|\approx 100$~MeV for $T=180$~MeV, and it is therefore only plotted up to this value. We also note that the thermal quark mass at zero momentum monotonically increases with temperature and that the splitting between the thermal quark mode and the plasmino mode decreases. The small kinks visible in the dispersion relations usually appear when the corresponding mode enters a continuum regime or passes a decay threshold in the spectral function.

As for the phonino mode, we first note its space-like character, with an energy always smaller than its spatial momentum, $\omega<|\vec{p}|$, which is evident from the light-cone indicated as long-dashed line at $\omega=|\vec{p}|$ in Fig.~\ref{fig:dispersion}. We also observe that the phonino branch continuously moves to smaller energies with increasing temperature until it switches roles with the anti-phonino mode at about $T=210$~MeV. At $T=300$~MeV we therefore only see the anti-phonino mode in Fig.~\ref{fig:dispersion} while the phonino mode is given by its reflection to negative energies. For the slope of the phonino branch we find
\begin{align}
\frac{\partial \omega_{\text{phonino}}}{{\partial |\vec{p}|}}\Big|_{|\vec{p}|=0, T=300~\text{MeV}}\approx \begin{cases}1/\sqrt{3} &\text{FRG}\\
1/3 &\text{1-loop} \end{cases}\,,
\end{align}
at $T=300$~MeV. We note that a value of $1/3$ is expected from analytical beyond-HTL calculations at high temperatures, see e.g.~Ref.\ \cite{BlaizotSatow2014}, while a value of $1/\sqrt{3}$ was for example found numerically in Ref.\ \cite{SuTywoniuk2015}. We also note that the sound velocity in an ideal fluid of massless particles is given by $c_s=1/\sqrt{3}$, owing to the well-known relation between energy density and pressure, $\epsilon=3p$. One might therefore expect that a value of $1/\sqrt{3}$ should be approached at high temperatures, which appears to be the case in the FRG calculation already at about $300$ MeV.

In general, however, we find that the slope of the phonino mode rather strongly depends on the temperature, cf.\ e.g.~the bottom and top panel of Fig.~\ref{fig:dispersion}, and that at higher temperatures, around $T=300$~MeV, it is not the phonino mode but the anti-phonino mode that is present at positive energies. This effect may be due to the non-trivial structure of the overall spectral function and its dependence on the temperature. In particular the close proximity to the thermal-continuum threshold seems to have a strong influence on the phonino and the anti-phonino branches. 

\section{Summary and Outlook}\label{sec:summary}

In this work we have studied quark spectral functions at finite temperature, finite chemical potential, and finite spatial momentum, in order to identify fermionic excitations in a hot and dense strongly interacting medium. As an effective model for the chiral aspects of QCD we have used the two-flavor quark-meson model. In order to include fluctuations we have employed the Functional Renormalization Group approach in the local potential approximation (LPA). The analytic continuation from imaginary to real energies was performed using the aFRG method, which allows to obtain analytically continued FRG flow equations for retarded two-point functions. The particular truncation scheme is thermodynamically consistent and preserves chiral symmetry and its explicit versus dynamical breaking patterns. As a simpler alternative scheme to calculate quark spectral functions we have also assessed an FRG-improved one-loop setup, which uses momentum-dependent quasi-particle masses extracted from the FRG calculation. A detailed comparison between the full FRG treatment and the FRG-improved one-loop setup shows quite remarkable agreement, revealing that the main effect of the fluctuations included in the FRG calculation can be reproduced by a suitably modified one-loop calculation. This offers a simple physical interpretation of the results and paves the way for systematic improvements in terms of a loop expansion.
 
We find that the quark spectral functions exhibit non-trivial in-medium modifications due to the influence of various decay and scattering channels. In particular, we were able to identify three different fermionic excitations: the thermal quark, the plasmino, and the phonino. We presented results on the dispersion relations of these collective excitations at finite temperature in the chiral limit. The thermal quark and the plasmino mode behave as expected from standard HTL calculations. The phonino, on the other hand, shows a strongly temperature-dependent dispersion relation, which eventually leads to a switching of the phonino and the anti-phonino mode at high temperatures.

The results presented in this work open up possibilities for several future applications. A direct application of the momentum-dependent in-medium spectral functions is to use them as input for the calculation of other real-time quantities such as transport coefficients. Another interesting possibility is given by improving the current truncation towards a self-consistent solution of the spectral function, which couples back to the effective potential, or to include higher orders in the derivative expansion of the effective action. Also, replacing the quarks by nucleon fields and their parity partners would allow to study the corresponding baryonic spectral functions in the parity-doublet model with fluctuations beyond mean-field as in Ref.\ \cite{Weyrich:2015hha}, in order to describe the liquid-gas transition of nuclear matter as well as the chiral transition at high baryon density in a unified framework. This can then furthermore be extended to include vector and axial-vector mesons along the lines of Ref.\ \cite{JungRenneckeTripoltEtAl2017} and study their spectral changes in dense nuclear matter.

\acknowledgments
This work was supported by the Deutsche Forschungsgemeinschaft (DFG) through the grant CRC-TR 211 ``Strong-interaction matter under extreme conditions'' and the German Federal Ministry of Education and Research (BMBF) through grants No.~05P18RFFCA and No.~05P18RGFCA. We also acknowledge the Goethe-HLR high-performance computing cluster, where some of the numerical calculations were performed.

\appendix

\section{Details of the FRG setup}
\label{app:FRG}

The three-dimensional bosonic and fermionic regulator functions are given by
\begin{align}
R^B_{k}(\vec q)&=(k^2-\vec q^{\,2})\theta(k^2-\vec q^{\,2})\,,\\
R^F_{k}(\vec q)&=i \slashed{\vec q} (\sqrt{k^2/\vec q^{\,2}}-1)
\theta(k^2-\vec q^{\,2})\,.
\end{align}

While the FRG flow for the effective average action explicitly contains the regulator $R_k$, physics at $k\to 0$ should not depend on a particular choice. For an up-to-date discussion of how to devise optimized regulators in a particular truncation where this can be quite non-trivial, see Ref.\ \cite{Pawlowski:2015mlf}.

The threshold functions appearing in Eq.~(\ref{eq:flow_pot}) are given by
\begin{align}
I_{k,\alpha}& =
\frac{k}{E_{k,\alpha}}\left[1+2\,n_B(E_{k,\alpha})\right]\, ,\\
I_{k,\psi}& =
\frac{k}{E_{k,\psi}}\left[1-n_F(E_{k,\psi}-\mu)-n_F(E_{k,\psi}+\mu)\right]\, ,
\end{align}
with the effective energies
\begin{align}
\label{eq:energies}
E_{k,\alpha}=\sqrt{k^2+m_{k,\alpha}^2}, \qquad \alpha \in \{\pi,\sigma,\psi\}\,,
\end{align}
and the masses
\begin{align}
\label{eq:masses}
m_{k,\pi}^2=2U_k',\quad m_{k,\sigma}^2=2U_k'+4 U_k''\phi^2,\quad m_{k,\psi}^2=h^2\phi^2\,.
\end{align}
The three-point vertex functions appearing in Eq.~(\ref{eq:flow_gamma2}) are given by
\begin{align}
\Gamma^{(3)}_{\bar \psi \psi  \phi_i}=h \begin{cases}1 &\text{for}\; i=0\\
i \gamma^5\tau^i &\text{for}\; i=1,2,3 \end{cases}\,.
\end{align}


The dressing functions defined in Eq.~(\ref{eq:Gamma2_R}) can be obtained from the full two-point function as follows,
\begin{align}
A_{k}(\omega,\vec{p}) &= - \frac{1}{4} \, \tr \Big( i\vec{\gamma}\hat{p}\: \Gamma^{(2)}_{k,\psi}(\omega,\vec{p})\Big)\, ,\\
B_{k}(\omega,\vec{p}) &= - \frac{1}{4} \, \tr \Big( \Gamma^{(2)}_{k,\psi}(\omega,\vec{p})\Big)\, ,\\
C_{k}(\omega,\vec{p}) &= \frac{1}{4} \, \tr \Big( \gamma_0 \Gamma^{(2)}_{k,\psi}(\omega,\vec{p})\Big)\, .
\end{align}
The flow equations for the individual dressing functions are then given by
\begin{align}
\label{eq:flow_eq_2PF}
\partial_kX_k(\omega,\vec{p})&=
\mathcal{J}^{(X)}_{k,\sigma\psi}(\omega,\vec{p})+
\mathcal{J}^{(X)}_{k,\psi\sigma}(\omega,\vec{p})\nonumber\\
&\quad \: \: \: +
3\,\mathcal{J}^{(X)}_{k,\pi\psi}(\omega,\vec{p})+
3\,\mathcal{J}^{(X)}_{k,\psi\pi}(\omega,\vec{p})\, ,
\end{align}
with $X\in \{A,B,C\}$ and the generalized loop functions
\begin{align}
\mathcal{J}^{(X)}_{k,\alpha\beta}(\omega,\vec{p})&=\int_{|\vec{q}\pm\vec{p}|\leq k }\frac{d^3q}{(2\pi)^3}\,
J^{(X)}_{k,\alpha\beta}(\omega,\vec{p},\vec{q}),
\end{align}
with $\alpha,\beta\in \{\sigma,\pi,\psi\}$, $X\in \{A,B,C\}$ and
\begin{equation}
\pm=
\begin{cases}
+&\text{for}\,\, \beta=\psi \\
-&\text{for}\,\, \alpha=\psi 
\end{cases}.
\end{equation}
The momentum integration therein is most conveniently performed using spherical coordinates,
\begin{align}
\int\frac{d^3q}{(2\pi)^3}=
\frac{1}{(2\pi)^3}\int d q_r \: q_r^2
\int d\theta\: \sin\theta
\int d\phi,
\end{align}
where the $\phi$-integration is trivial and the $\theta$-integration can be performed analytically.

In App.~\ref{app:loop_functions} we provide explicit expressions for the loop functions $J^{(X)}_{k,\alpha\beta}(\omega,\vec{p},\vec{q})$, where we use the following notation. The function $F(\theta)$ is given by
\begin{equation}
F(\theta)=
\begin{cases}
\frac{p_z+q_r \cos \theta}{\sqrt{p_z^2+q_r^2+2p_zq_r\cos \theta}}&\text{for}\,\, |\vec{q}|>k \\
\cos\theta&\text{for}\,\, |\vec{q}|\leq k
\end{cases}.
\end{equation}
The non-regulated energies $\tilde{E}_\alpha$ are given by
\begin{align}
\tilde{E}_\alpha=\sqrt{q_r^2+m_\alpha^2}, \qquad \alpha \in \{\pi,\sigma,\psi\}\,,
\end{align}
where $q_r$ is substituted by $k$ for $|\vec{q}|\leq k$, also in the expressions for the loop functions. Moreover, we use the following short-hand notation,
\begin{align}
\omega&=\omega + i \epsilon + \mu,\\
n_B&=n_B(E_\alpha),\\
\tilde{n}_B&=n_B(\tilde{E}_\alpha),\\
n_F^\pm&=n_F(E_\psi\pm\mu),\\
\tilde{n}_F^\pm&=n_F(\tilde{E}_\psi\pm\mu),
\end{align}
and
\begin{equation}
s=
\begin{cases}
+&\text{for}\,\, \alpha=\sigma \\
-&\text{for}\,\, \alpha=\pi
\end{cases}.
\end{equation}
We also note that the limit $\epsilon\rightarrow 0$, which is implied implicitly, can be performed analytically for the flow equation of the imaginary part of the two-point functions. This can be seen by rewriting the imaginary part of the loop functions by using the Dirac-Sokhotsky identities,
\begin{align}
\lim_{\epsilon \rightarrow 0}\text{Im}\frac{1}{\omega+ i\epsilon \pm E_\alpha\pm E_\beta}\rightarrow -\pi\delta(\omega\pm E_\alpha \pm E_\beta),\\
\lim_{\epsilon \rightarrow
	0}\text{Im}\frac{1}{(\omega+ i\epsilon \pm E_\alpha\pm E_\beta)^2}\rightarrow \pi\delta'(\omega \pm E_\alpha\pm E_\beta).
\end{align}
The flow equation for the imaginary part of the retarded two-point function then reduces to a sum over a few values $k_0$ that correspond to the scales where one of the arguments of the delta function becomes zero, see Ref.\ \cite{JungRenneckeTripoltEtAl2017} for details.




Based on the solution of the flow equations of the quark two-point function, the quark and anti-quark spectral function at zero momentum are then given by
\begin{align}
\rho_{k,L}^\pm(\omega)&=\frac{1}{\pi}
\frac{\Im C_k\mp\Im B_k}{(\Re C_k\mp\Re B_k)^2+(\Im C_k\mp\Im B_k)^2}\, .
\end{align}
Similarly, the quark and anti-quark spectral functions at zero quark mass are given by
\begin{align}
\rho_{k,P}^\pm (\omega,\vec{p})&=\frac{1}{\pi}
\frac{\Im C_k\mp\Im A_k}{(\Re C_k\mp\Re A_k)^2+(\Im C_k\mp\Im A_k)^2}\, .
\end{align}
These expressions can be obtained by applying the corresponding projection operators to the quark spectral function, 
\begin{align}
\rho_{k,L}^\pm(\omega) 
&= \frac{1}{2}\text{Tr}[\rho_{k,\psi}(\omega,0)\gamma_0 L^\pm] \\
&=\rho^{(C)}_{k,\psi}(\omega)\pm\rho^{(B)}_{k,\psi}(\omega),
\end{align}
and
\begin{align}
\rho_{k,P}^\pm(\omega,\vec{p})
&= \frac{1}{2}\text{Tr}[\rho_{k,\psi}(\omega,\vec{p})\gamma_0 P^\pm]  \\
&=\rho^{(C)}_{k,\psi}(\omega,\vec{p})\pm\rho^{(A)}_{k,\psi}(\omega,\vec{p}).
\end{align}
We note that in general $\rho_{k,L}^\pm(\omega)$ and $\rho_{k,P}^\pm(\omega,\vec{p})$ are neither even nor odd functions. Instead, charge-conjugation symmetry requires
\begin{align}
\rho_{k,L}^\pm(\omega)&=\rho_{k,L}^\mp(-\omega),\label{eq:rel1}\\
\rho_{k,P}^\pm(\omega,\vec{p})&=\rho_{k,P}^\mp(-\omega,\vec{p}),\label{eq:rel2}
\end{align}
see for example \cite{KarschKitazawa2009, Kaczmarek:2012mb}.



\section{Details of the 1-loop setup}
\label{app:1_loop}

The loop functions introduced in Eq.~(\ref{eq:one_loop_coeffs}) are defined as
\begin{align}
\mathcal{L}^{(X)}_{\alpha\beta}(\omega,\vec{p})=\int\frac{d^3q}{(2\pi)^3}L^{(X)}_{\alpha\beta}(\omega,\vec{p},\vec{q}),
\end{align}
with $\alpha,\beta\in \{\sigma,\pi,\psi\}$ and $X\in \{A,B,C\}$. The momentum integration is again performed using spherical coordinates,
\begin{align}
\int\frac{d^3q}{(2\pi)^3}=
\frac{1}{(2\pi)^3}\int_{0}^{\Lambda} d q_r \: q_r^2
\int_{0}^{\pi}d\theta\: \sin\theta
\int_{0}^{2\pi}d\phi,
\end{align}
where $\Lambda=1$~GeV is used as UV-cutoff, as for the FRG calculation. 

In App.~\ref{app:loop_functions}, we provide explicit expressions for the loop functions $L^{(X)}_{\alpha\beta}(\omega,\vec{p},\vec{q})$, where we use the following notation. The energies are defined as 
\begin{align}
E_\alpha&=\sqrt{m_\alpha^2+q_r^2},\\
E_\psi&=\sqrt{m_\psi^2+q_r^2},\\
\tilde{E}_\alpha&=\sqrt{m_\alpha^2+q_r^2+p_z^2-2 q_r p_z \cos\theta},\\
\tilde{E}_\psi&=\sqrt{m_\psi^2+q_r^2+p_z^2+2 q_r p_z \cos\theta}
\end{align}
with $\alpha \in \{\pi,\sigma\}$, $m_\psi=h\sigma_0$, and the momentum-dependent meson masses are taken to be the scale-dependent masses from the FRG calculation, $m_\alpha^2(q_r)=m_\alpha^2(k)$.

\section{Relation between FRG and 1-loop setup}
\label{app:relation}

We note that the one-loop expressions for $\Delta X(\omega,\vec{p})$, as defined in Sec.~\ref{sec:1_loop}, are closely connected to the corresponding FRG expressions. To see this, we first define
\begin{align}
\Delta X^{\text{1-loop}}(\omega)&=\frac{1}{2\pi^2}\int_0^\Lambda dk\: k^2 \: \sum_{\alpha\beta} L_{\alpha\beta}^{(X)}(\omega),\\
\Delta X^{\text{FRG}}(\omega)&=\frac{1}{2\pi^2}\int_0^\Lambda dk\: \frac{k^3}{3} \: \sum_{\alpha\beta} J_{\alpha\beta}^{(X)}(\omega),
\end{align}
where $\vec{p}=0$, $X\in \{B,C\}$, $q_r$ was replaced by $k$ in the 1-loop expression, and the sum is over all loop functions. Noting that the loop functions for $X\in \{B,C\}$ are related by
\begin{align}
\sum \partial_k L^{(X)}(\omega)=-\sum J^{(X)}(\omega),
\end{align}
we find, by using integration by parts, that the two expressions for $\Delta X$ only differ by a boundary term,
\begin{align}
\Delta X^{\text{1-loop}}(\omega)=\Delta X^{\text{FRG}}(\omega)+\frac{1}{2\pi^2}\frac{k^3}{3}\sum L^{(X)}(\omega)\Big|^{\Lambda}_{0}.
\end{align}

\begin{figure}[t!]
\includegraphics[width=\columnwidth]{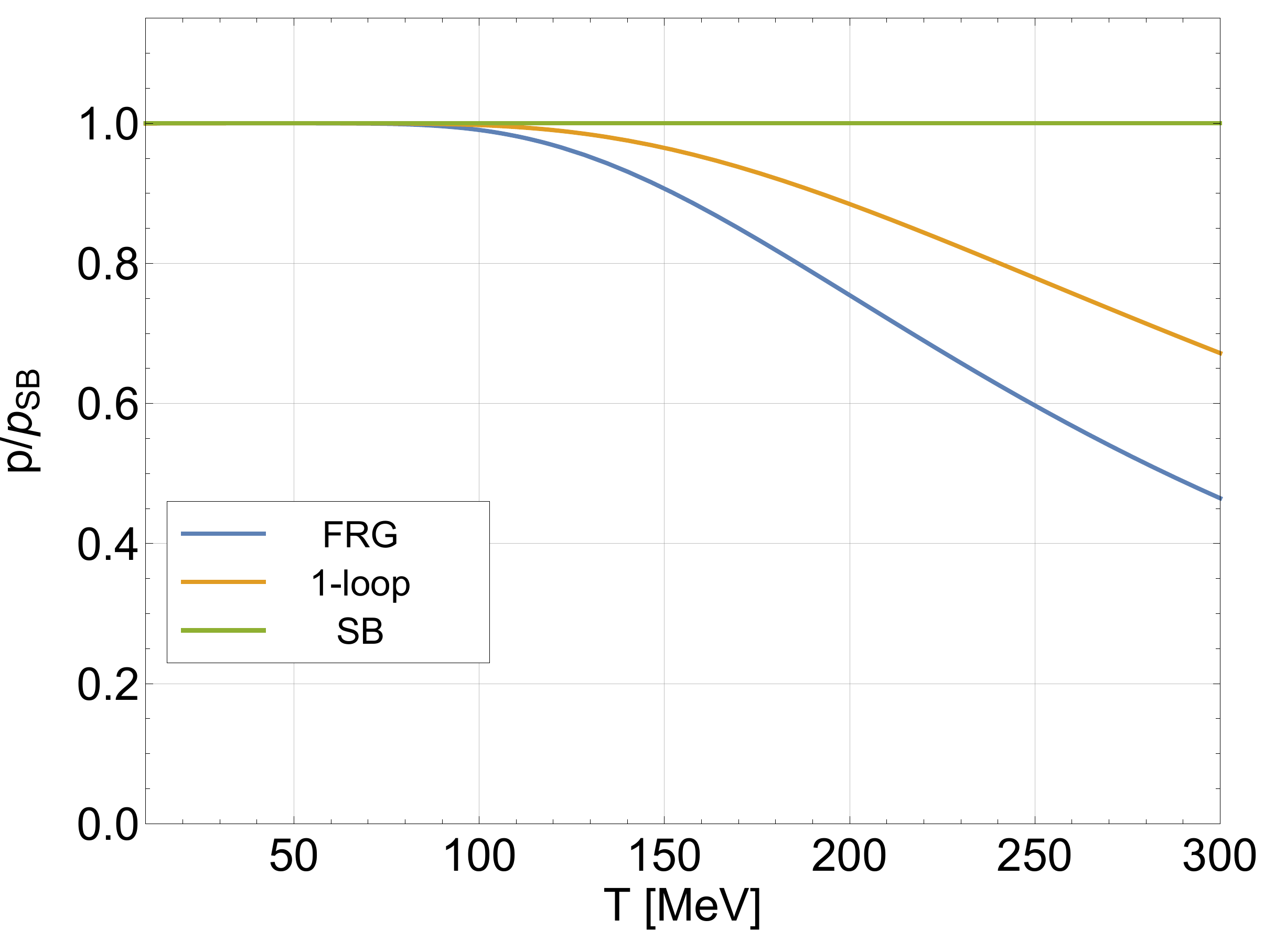}\\[-2mm]
\caption{Comparison of the pressure as obtained from the FRG setup to the one from the one-loop calculation, for a massless system of non-interacting bosons and a UV cutoff of $\Lambda=1$~GeV, normalized to the corresponding Stefan-Boltzmann value $p_\text{SB}=\pi^2T^4/90$, which corresponds to $\Lambda\rightarrow\infty$.}
\label{fig:comparison}
\end{figure}

A similar relation can be obtained for the effective potential. Following the standard derivation for the thermodynamic potential of a non-interacting system of bosons and fermions in thermal field theory, see for example Ref.\ \cite{Lebellac:1996}, we find
\begin{align}
U^{\text{1-loop}}=\int_0^\Lambda\frac{dk}{4\pi^2}k^2\left(K_{k,\sigma}+3K_{k,\pi} -4N_c N_f K_{k,\psi}\right),
\end{align}
where we defined
\begin{align}
K_{k,\alpha}&=E_{k,\alpha}+2T\log \left[1-\exp(-E_{k,\alpha}/T)\right] ,\\
K_{k,\psi}&=E_{k,\psi}+T\log\left\{1+\exp\left[-(E_{k,\psi}-\mu)/T\right]\right\}\nonumber\\
&\hspace{14mm}+T\log\left\{1+\exp\left[-(E_{k,\psi}+\mu)/T\right]\right\},
\end{align}
with $\alpha\in \{\sigma,\pi\}$, see also Ref.\ \cite{Strodthoff:2011tz}. This expression needs to be compared to the infrared value of the effective potential as obtained from the FRG setup,
\begin{align}
U^{\text{FRG}}_{\text{IR}}=U^{\text{FRG}}_{\text{UV}}-\Delta U^{\text{FRG}},
\end{align}
where we will set $U^{\text{FRG}}_{\text{UV}}=0$ for ease of comparison. We then have
\begin{align}
U^{\text{FRG}}_{\text{IR}}=-\int_0^\Lambda \frac{dk}{4\pi^2} \frac{k^3}{3}\left( I_{k,\sigma} + 3 I_{k,\pi} -4\Nc \Nf I_{k,\psi}\right),
\end{align}
cf.~Eq.~(\ref{eq:flow_pot}). We note that the loop functions $K$ and $I$ fulfill a similar relation as $L$ and $J$ for the two-point function, namely
\begin{align}
I_{k,\alpha}=\partial_k K_{k,\alpha}
\end{align}
for $\alpha\in \{\sigma,\pi,\psi\}$. We then find, by using integration by parts again, that the two expressions for the effective potential only differ by a boundary term,
\begin{align}
\label{eq:app1}
U^{\text{1-loop}}=U^{\text{FRG}}_{\text{IR}}+\frac{1}{4\pi^2}\frac{k^3}{3}\sum_\alpha K_{k,\alpha}\Big|^{\Lambda}_{0}.
\end{align}
We will now normalize the potential to its value at $T=0$ and compare the results for the pressure as a function of temperature,
\begin{align}
-p=U(T)-U(T=0).
\end{align}
In Fig.~\ref{fig:comparison} we compare the pressure as obtained from the FRG setup to the one from the one-loop calculation, for a massless system of non-interacting bosons, normalized to the Stefan-Boltzmann value of
\begin{align}
p_\text{SB}=\frac{\pi^2T^4}{90},
\end{align}
which corresponds to $\Lambda\rightarrow\infty$.

We find that the agreement between the FRG and the one-loop result is very good at low temperatures, while at higher temperatures the finite UV cutoff leads to discrepancies between the two frameworks and to an overall reduction of the pressure as compared to the exact result. For $\Lambda\rightarrow\infty$ the boundary term in Eq.~(\ref{eq:app1}) vanishes, after subtracting the vacuum contribution, and we have
\begin{align}
p^{\text{FRG}}_{\Lambda\rightarrow\infty}=p^{\text{1-loop}}_{\Lambda\rightarrow\infty}.
\end{align}

We note here that if the temperature becomes too large compared to the UV cutoff scale, in our case of ${T_{{\rm max}} \approx \Lambda/2\pi \approx 170\,{\rm MeV}}$, the assumption of a temperature-independent effective action in the UV breaks down. If one wants to extend the accessible temperature range for a fixed UV cutoff, the results have to be supplemented by perturbative input \cite{Braun:2003ii,Herbst:2010rf,StrodthoffSmekal2014}.

\section{Loop functions}
\label{app:loop_functions}

In this appendix we give explicit expressions for the various loop functions.

\begin{align}
&J^{(A)}_{k,\alpha\psi}(\omega,\vec{p},\vec{q})=\nonumber
\\
&+\frac{1}{(\omega+E_\alpha+\tilde{E}_\psi)}
\frac{ -kq_rh^2\cos\theta}{4E_\alpha^3\tilde{E}_\psi}
\Big(
1+n_B-\tilde{n}_F^+
-E_\alpha n'_B
\Big)
\nonumber
\\
&+\frac{1}{(\omega+E_\alpha+\tilde{E}_\psi)^2}
\frac{ -kq_rh^2\cos\theta}{4E_\alpha^2\tilde{E}_\psi}
\Big(
1+n_B-\tilde{n}_F^+
\Big)
\nonumber
\\
&+\frac{1}{(\omega-E_\alpha-\tilde{E}_\psi)}
\frac{ kq_rh^2\cos\theta}{4E_\alpha^3\tilde{E}_\psi}
\Big(
1+n_B-\tilde{n}_F^-
-E_\alpha n'_B
\Big)
\nonumber
\\
&+\frac{1}{(\omega-E_\alpha-\tilde{E}_\psi)^2}
\frac{ -kq_rh^2\cos\theta}{4E_\alpha^2\tilde{E}_\psi}
\Big(
1+n_B-\tilde{n}_F^-
\Big)
\nonumber
\\
&+\frac{1}{(\omega+E_\alpha-\tilde{E}_\psi)}
\frac{ kq_rh^2\cos\theta}{4E_\alpha^3\tilde{E}_\psi}
\Big(
n_B+\tilde{n}_F^-
-E_\alpha n'_B
\Big)
\nonumber
\\
&+\frac{1}{(\omega+E_\alpha-\tilde{E}_\psi)^2}
\frac{ kq_rh^2\cos\theta}{4E_\alpha^2\tilde{E}_\psi}
\Big(
n_B+\tilde{n}_F^-
\Big)
\nonumber
\\
&+\frac{1}{(\omega-E_\alpha+\tilde{E}_\psi)}
\frac{- kq_rh^2\cos\theta}{4E_\alpha^3\tilde{E}_\psi}
\Big(
n_B+\tilde{n}_F^+
-E_\alpha n'_B
\Big)
\nonumber
\\
&+\frac{1}{(\omega-E_\alpha+\tilde{E}_\psi)^2}
\frac{ kq_rh^2\cos\theta}{4E_\alpha^2\tilde{E}_\psi}
\Big(
n_B+\tilde{n}_F^+
\Big)
\label{eq:JA1}
\end{align}

\begin{align}
&J^{(B)}_{k,\alpha\psi}(\omega,\vec{p},\vec{q})=\nonumber
\\
&+\frac{1}{(\omega+E_\alpha+\tilde{E}_\psi)}
\frac{s kh^2m_\psi}{4E_\alpha^3\tilde{E}_\psi}
\Big(
1+n_B-\tilde{n}_F^+
-E_\alpha n'_B
\Big)
\nonumber
\\
&+\frac{1}{(\omega+E_\alpha+\tilde{E}_\psi)^2}
\frac{s kh^2m_\psi}{4E_\alpha^2\tilde{E}_\psi}
\Big(
1+n_B-\tilde{n}_F^+
\Big)
\nonumber
\\
&+\frac{1}{(\omega-E_\alpha-\tilde{E}_\psi)}
\frac{-s kh^2m_\psi}{4E_\alpha^3\tilde{E}_\psi}
\Big(
1+n_B-\tilde{n}_F^-
-E_\alpha n'_B
\Big)
\nonumber
\\
&+\frac{1}{(\omega-E_\alpha-\tilde{E}_\psi)^2}
\frac{s kh^2m_\psi}{4E_\alpha^2\tilde{E}_\psi}
\Big(
1+n_B-\tilde{n}_F^-
\Big)
\nonumber
\\
&+\frac{1}{(\omega+E_\alpha-\tilde{E}_\psi)}
\frac{-s kh^2m_\psi}{4E_\alpha^3\tilde{E}_\psi}
\Big(
n_B+\tilde{n}_F^-
-E_\alpha n'_B
\Big)
\nonumber
\\
&+\frac{1}{(\omega+E_\alpha-\tilde{E}_\psi)^2}
\frac{-s kh^2m_\psi}{4E_\alpha^2\tilde{E}_\psi}
\Big(
n_B+\tilde{n}_F^-
\Big)
\nonumber
\\
&+\frac{1}{(\omega-E_\alpha+\tilde{E}_\psi)}
\frac{s kh^2m_\psi}{4E_\alpha^3\tilde{E}_\psi}
\Big(
n_B+\tilde{n}_F^+
-E_\alpha n'_B
\Big)
\nonumber
\\
&+\frac{1}{(\omega-E_\alpha+\tilde{E}_\psi)^2}
\frac{-s kh^2m_\psi}{4E_\alpha^2\tilde{E}_\psi}
\Big(
n_B+\tilde{n}_F^+
\Big)
\end{align}

\begin{align}
&J^{(C)}_{k,\alpha\psi}(\omega,\vec{p},\vec{q})=\nonumber\\
&+\frac{1}{(\omega+E_\alpha+\tilde{E}_\psi+\mu)}
\frac{ kh^2}{4E_\alpha^3}
\Big(
1+n_B-\tilde{n}_F^+
-E_\alpha n'_B
\Big)
\nonumber
\\
&+\frac{1}{(\omega+E_\alpha+\tilde{E}_\psi+\mu)^2}
\frac{kh^2}{4E_\alpha^2}
\Big(
1+n_B-\tilde{n}_F^+
\Big)
\nonumber
\\
&+\frac{1}{(\omega-E_\alpha-\tilde{E}_\psi+\mu)}
\frac{kh^2}{4E_\alpha^3}
\Big(
1+n_B-\tilde{n}_F^-
-E_\alpha n'_B
\Big)
\nonumber
\\
&+\frac{1}{(\omega-E_\alpha-\tilde{E}_\psi+\mu)^2}
\frac{-kh^2}{4E_\alpha^2}
\Big(
1+n_B-\tilde{n}_F^-
\Big)
\nonumber
\\
&+\frac{1}{(\omega+E_\alpha-\tilde{E}_\psi+\mu)}
\frac{kh^2}{4E_\alpha^3}
\Big(
n_B+\tilde{n}_F^-
-E_\alpha n'_B
\Big)
\nonumber
\\
&+\frac{1}{(\omega+E_\alpha-\tilde{E}_\psi+\mu)^2}
\frac{kh^2}{4E_\alpha^2}
\Big(
n_B+\tilde{n}_F^-
\Big)
\nonumber
\\
&+\frac{1}{(\omega-E_\alpha+\tilde{E}_\psi+\mu)}
\frac{kh^2}{4E_\alpha^3}
\Big(
n_B+\tilde{n}_F^+
-E_\alpha n'_B
\Big)
\nonumber
\\
&+\frac{1}{(\omega-E_\alpha+\tilde{E}_\psi+\mu)^2}
\frac{ -kh^2}{4E_\alpha^2}
\Big(
n_B+\tilde{n}_F^+
\Big)
\end{align}

\begin{align}
&J^{(A)}_{k,\psi\alpha}(\omega,\vec{p},\vec{q})=\nonumber\\
&+\frac{1}{(\omega+E_\psi+\tilde{E}_\alpha)}
\frac{h^2m_\psi^2F(\theta)}{4E_\psi^3\tilde{E}_\alpha}
\Big(
1+\tilde{n}_B-n_F^+
-\frac{E_\psi k^2}{m_\psi^2} n_F^{\prime +}
\Big)
\nonumber
\\
&+\frac{1}{(\omega+E_\psi+\tilde{E}_\alpha)^2}
\frac{ -k^2h^2 F(\theta)}{4E_\psi^2\tilde{E}_\alpha}
\Big(
1+\tilde{n}_B-n_F^+
\Big)
\nonumber
\\
&+\frac{1}{(\omega-E_\psi-\tilde{E}_\alpha)}
\frac{ -h^2m_\psi^2 F(\theta)}{4E_\psi^3\tilde{E}_\alpha}
\Big(
1+\tilde{n}_B-n_F^-
-\frac{E_\psi k^2}{m_\psi^2} n_F^{\prime -}
\Big)
\nonumber
\\
&+\frac{1}{(\omega-E_\psi-\tilde{E}_\alpha)^2}
\frac{- k^2h^2 F(\theta)}{4E_\psi^2\tilde{E}_\alpha}
\Big(
1+\tilde{n}_B-n_F^-
\Big)
\nonumber
\\
&+\frac{1}{(\omega+E_\psi-\tilde{E}_\alpha)}
\frac{h^2m_\psi^2 F(\theta)}{4E_\psi^3\tilde{E}_\alpha}
\Big(
n_F^+ +\tilde{n}_B
+\frac{E_\psi k^2}{m_\psi^2} n_F^{\prime +}
\Big)
\nonumber
\\
&+\frac{1}{(\omega+E_\psi-\tilde{E}_\alpha)^2}
\frac{-k^2h^2 F(\theta)}{4E_\psi^2\tilde{E}_\alpha}
\Big(
n_F^+ +\tilde{n}_B
\Big)
\nonumber
\\
&+\frac{1}{(\omega-E_\psi+\tilde{E}_\alpha)}
\frac{-h^2m_\psi^2 F(\theta)}{4E_\psi^3\tilde{E}_\alpha}
\Big(
n_F^- +\tilde{n}_B
+\frac{E_\psi k^2}{m_\psi^2} n_F^{\prime -}
\Big)
\nonumber
\\
&+\frac{1}{(\omega-E_\psi+\tilde{E}_\alpha)^2}
\frac{-k^2h^2 F(\theta)}{4E_\psi^2\tilde{E}_\alpha}
\Big(
n_F^- +\tilde{n}_B
\Big)
\end{align}

\begin{align}
&J^{(B)}_{k,\psi\alpha}(\omega,\vec{p},\vec{q})=\nonumber\\
&+\frac{1}{(\omega+E_\psi+\tilde{E}_\alpha)}
\frac{s kh^2m_\psi}{4E_\psi^3\tilde{E}_\alpha}
\Big(
1+\tilde{n}_B-n_F^+
+E_\psi n_F^{\prime +}
\Big)
\nonumber
\\
&+\frac{1}{(\omega+E_\psi+\tilde{E}_\alpha)^2}
\frac{s kh^2m_\psi}{4E_\psi^2\tilde{E}_\alpha}
\Big(
1+\tilde{n}_B-n_F^+
\Big)
\nonumber
\\
&+\frac{1}{(\omega-E_\psi-\tilde{E}_\alpha)}
\frac{-s kh^2m_\psi}{4E_\psi^3\tilde{E}_\alpha}
\Big(
1+\tilde{n}_B-n_F^-
+E_\psi n_F^{\prime -}
\Big)
\nonumber
\\
&+\frac{1}{(\omega-E_\psi-\tilde{E}_\alpha)^2}
\frac{s kh^2m_\psi}{4E_\psi^2\tilde{E}_\alpha}
\Big(
1+\tilde{n}_B-n_F^-
\Big)
\nonumber
\\
&+\frac{1}{(\omega+E_\psi-\tilde{E}_\alpha)}
\frac{s kh^2m_\psi}{4E_\psi^3\tilde{E}_\alpha}
\Big(
n_F^+ +\tilde{n}_B
-E_\psi n_F^{\prime +}
\Big)
\nonumber
\\
&+\frac{1}{(\omega+E_\psi-\tilde{E}_\alpha)^2}
\frac{s kh^2m_\psi}{4E_\psi^2\tilde{E}_\alpha}
\Big(
n_F^+ +\tilde{n}_B
\Big)
\nonumber
\\
&+\frac{1}{(\omega-E_\psi+\tilde{E}_\alpha)}
\frac{-s kh^2m_\psi}{4E_\psi^3\tilde{E}_\alpha}
\Big(
n_F^- +\tilde{n}_B
-E_\psi n_F^{\prime -}
\Big)
\nonumber
\\
&+\frac{1}{(\omega-E_\psi+\tilde{E}_\alpha)^2}
\frac{s kh^2m_\psi}{4E_\psi^2\tilde{E}_\alpha}
\Big(
n_F^- +\tilde{n}_B
\Big)
\end{align}

\begin{align}
&J^{(C)}_{k,\psi\alpha}(\omega,\vec{p},\vec{q})=\nonumber\\
&+\frac{1}{(\omega+E_\psi+\tilde{E}_\alpha)}
\frac{ kh^2}{4E_\psi \tilde{E}_\alpha}
\Big(
n_F^{\prime +}
\Big)
\nonumber
\\
&+\frac{1}{(\omega+E_\psi+\tilde{E}_\alpha)^2}
\frac{ kh^2}{4E_\psi \tilde{E}_\alpha}
\Big(
1+\tilde{n}_B-n_F^+
\Big)
\nonumber
\\
&+\frac{1}{(\omega-E_\psi-\tilde{E}_\alpha)}
\frac{ kh^2}{4E_\psi \tilde{E}_\alpha}
\Big(
n_F^{\prime -}
\Big)
\nonumber
\\
&+\frac{1}{(\omega-E_\psi-\tilde{E}_\alpha)^2}
\frac{- kh^2}{4E_\psi \tilde{E}_\alpha}
\Big(
1+\tilde{n}_B-n_F^-
\Big)
\nonumber
\\
&+\frac{1}{(\omega+E_\psi-\tilde{E}_\alpha)}
\frac{ -kh^2}{4E_\psi \tilde{E}_\alpha}
\Big(
n_F^{\prime +}
\Big)
\nonumber
\\
&+\frac{1}{(\omega+E_\psi-\tilde{E}_\alpha)^2}
\frac{ kh^2}{4E_\psi \tilde{E}_\alpha}
\Big(
n_F^+ +\tilde{n}_B
\Big)
\nonumber
\\
&+\frac{1}{(\omega-E_\psi+\tilde{E}_\alpha)}
\frac{- kh^2}{4E_\psi \tilde{E}_\alpha}
\Big(
n_F^{\prime -}
\Big)
\nonumber
\\
&+\frac{1}{(\omega-E_\psi+\tilde{E}_\alpha)^2}
\frac{-kh^2}{4E_\psi \tilde{E}_\alpha}
\Big(
n_F^- +\tilde{n}_B
\Big)
\label{eq:JC2}
\end{align}



\begin{align}
&L^{(A)}_{\alpha\psi}(\omega,\vec{p},\vec{q})=\nonumber
\\
&+\frac{1}{(\omega+E_\alpha+\tilde{E}_\psi)}
\frac{-h^2(p_z+q_r\cos\theta)}{8E_\alpha\tilde{E}_\psi}
\Big(
1+n_B-\tilde{n}_F^+
\Big)
\nonumber
\\
&+\frac{1}{(\omega-E_\alpha-\tilde{E}_\psi)}
\frac{h^2(p_z+q_r\cos\theta)}{8E_\alpha\tilde{E}_\psi}
\Big(
1+n_B-\tilde{n}_F^-
\Big)
\nonumber
\\
&+\frac{1}{(\omega+E_\alpha-\tilde{E}_\psi)}
\frac{h^2(p_z+q_r\cos\theta)}{8E_\alpha\tilde{E}_\psi}
\Big(
n_B+\tilde{n}_F^-
\Big)
\nonumber
\\
&+\frac{1}{(\omega-E_\alpha+\tilde{E}_\psi)}
\frac{-h^2(p_z+q_r\cos\theta)}{8E_\alpha\tilde{E}_\psi}
\Big(
n_B+\tilde{n}_F^+
\Big)
\end{align}

\begin{align}
&L^{(B)}_{\alpha\psi}(\omega,\vec{p},\vec{q})=\nonumber
\\
&+\frac{1}{(\omega+E_\alpha+\tilde{E}_\psi)}
\frac{s h^2 m_\psi}{8E_\alpha\tilde{E}_\psi}
\Big(
1+n_B-\tilde{n}_F^+
\Big)
\nonumber
\\
&+\frac{1}{(\omega-E_\alpha-\tilde{E}_\psi)}
\frac{-s h^2 m_\psi}{8E_\alpha\tilde{E}_\psi}
\Big(
1+n_B-\tilde{n}_F^-
\Big)
\nonumber
\\
&+\frac{1}{(\omega+E_\alpha-\tilde{E}_\psi)}
\frac{-s h^2 m_\psi}{8E_\alpha\tilde{E}_\psi}
\Big(
n_B+\tilde{n}_F^-
\Big)
\nonumber
\\
&+\frac{1}{(\omega-E_\alpha+\tilde{E}_\psi)}
\frac{s h^2 m_\psi}{8E_\alpha\tilde{E}_\psi}
\Big(
n_B+\tilde{n}_F^+
\Big)
\end{align}

\begin{align}
&L^{(C)}_{\alpha\psi}(\omega,\vec{p},\vec{q})=\nonumber
\\
&+\frac{1}{(\omega+E_\alpha+\tilde{E}_\psi)}
\frac{h^2}{8E_\alpha}
\Big(
1+n_B-\tilde{n}_F^+
\Big)
\nonumber
\\
&+\frac{1}{(\omega-E_\alpha-\tilde{E}_\psi)}
\frac{h^2}{8E_\alpha}
\Big(
1+n_B-\tilde{n}_F^-
\Big)
\nonumber
\\
&+\frac{1}{(\omega+E_\alpha-\tilde{E}_\psi)}
\frac{h^2}{8E_\alpha}
\Big(
n_B+\tilde{n}_F^-
\Big)
\nonumber
\\
&+\frac{1}{(\omega-E_\alpha+\tilde{E}_\psi)}
\frac{h^2}{8E_\alpha}
\Big(
n_B+\tilde{n}_F^+
\Big)
\end{align}

\begin{align}
&L^{(A)}_{\psi\alpha}(\omega,\vec{p},\vec{q})=\nonumber
\\
&+\frac{1}{(\omega+E_\psi+\tilde{E}_\alpha)}
\frac{-h^2q_r\cos\theta}{8\tilde{E}_\alpha E_\psi}
\Big(
1+\tilde{n}_B-n_F^+
\Big)
\nonumber
\\
&+\frac{1}{(\omega-E_\psi-\tilde{E}_\alpha)}
\frac{h^2q_r\cos\theta}{8\tilde{E}_\alpha E_\psi}
\Big(
1+\tilde{n}_B-n_F^-
\Big)
\nonumber
\\
&+\frac{1}{(\omega+E_\psi-\tilde{E}_\alpha)}
\frac{-h^2q_r\cos\theta}{8\tilde{E}_\alpha E_\psi}
\Big(
\tilde{n}_B+n_F^+
\Big)
\nonumber
\\
&+\frac{1}{(\omega-E_\psi+\tilde{E}_\alpha)}
\frac{h^2q_r\cos\theta}{8\tilde{E}_\alpha E_\psi}
\Big(
\tilde{n}_B+n_F^-
\Big)
\end{align}

\begin{align}
&L^{(B)}_{\psi\alpha}(\omega,\vec{p},\vec{q})=\nonumber
\\
&+\frac{1}{(\omega+E_\psi+\tilde{E}_\alpha)}
\frac{s h^2 m_\psi}{8E_\psi\tilde{E}_\alpha}
\Big(
1+\tilde{n}_B-n_F^+
\Big)
\nonumber
\\
&+\frac{1}{(\omega-E_\psi-\tilde{E}_\alpha)}
\frac{-s h^2 m_\psi}{8E_\psi\tilde{E}_\alpha}
\Big(
1+\tilde{n}_B-n_F^-
\Big)
\nonumber
\\
&+\frac{1}{(\omega+E_\psi-\tilde{E}_\alpha)}
\frac{s h^2 m_\psi}{8E_\psi\tilde{E}_\alpha}
\Big(
\tilde{n}_B+n_F^+
\Big)
\nonumber
\\
&+\frac{1}{(\omega-E_\psi+\tilde{E}_\alpha)}
\frac{-s h^2 m_\psi}{8E_\psi\tilde{E}_\alpha}
\Big(
\tilde{n}_B+n_F^-
\Big)
\end{align}

\begin{align}
&L^{(C)}_{\psi\alpha}(\omega,\vec{p},\vec{q})=\nonumber
\\
&+\frac{1}{(\omega+E_\psi+\tilde{E}_\alpha)}
\frac{h^2}{8\tilde{E}_\alpha}
\Big(
1+\tilde{n}_B-n_F^+
\Big)
\nonumber
\\
&+\frac{1}{(\omega-E_\psi-\tilde{E}_\alpha)}
\frac{h^2}{8\tilde{E}_\alpha}
\Big(
1+\tilde{n}_B-n_F^-
\Big)
\nonumber
\\
&+\frac{1}{(\omega+E_\psi-\tilde{E}_\alpha)}
\frac{h^2}{8\tilde{E}_\alpha}
\Big(
\tilde{n}_B+n_F^+
\Big)
\nonumber
\\
&+\frac{1}{(\omega-E_\psi+\tilde{E}_\alpha)}
\frac{h^2}{8\tilde{E}_\alpha}
\Big(
\tilde{n}_B+n_F^-
\Big)
\end{align}


\end{document}